\documentclass{article}

\usepackage[version=3]{mhchem} % Package for chemical equation typesetting
\usepackage{siunitx} % Provides the \SI{}{} and \si{} command for typesetting SI units
\usepackage{graphicx} % Required for the inclusion of images
\usepackage{epsfig}
\usepackage{natbib} % Required to change bibliography style to APA
\usepackage{times}
\usepackage{bm}
\usepackage{natbib}
\usepackage{amsmath, amssymb,wasysym, pstricks}
\usepackage{amsfonts,amssymb,mathrsfs}
\usepackage[plain,noend]{algorithm2e}
\usepackage{tikz}
\usetikzlibrary{bayesnet}
\usepackage[margin=1in,footskip=0.25in]{geometry}

\usepackage{thmtools}
\usepackage{amsfonts}

\newtheorem{thm}{Theorem}[section]

\newtheorem{prop}{Proposition}[section]
\newtheorem{definition}{Definition}[section]
\newtheorem{example}{Example}[section]
\newenvironment{proof}{\paragraph{Proof:}}{\hfill$\square$}

\title{Evaluating and Optimizing Network Sampling Designs: Decision Theory and Information Theory Perspectives
\protect\thanks{Dr. Sim\'on Lunag\'omez is a Lecturer at the Department of Mathematics and Statistics at Lancaster University (s.lunagomez@lancaster.ac.uk). Marios Papamichalis is a Postdoctoral Researcher at the Statistics Department at Purdue University (mpapamic@purdue.edu). Dr. Patrick J. Wolfe is the Frederick L. Hovde Dean of the College of Science and Miller Family Professor of Statistics at Purdue University (patrick@purdue.edu). Dr. Edoardo M. Airoldi is the Millard E. Gladfelter Professor of Statistics and Data Science at Temple University (airoldi@temple.edu). He also serves as Director of the Fox School’s Data Science Center. }}
\author{Sim\'on Lunag\'omez, Marios Papamichalis, Patrick J. Wolfe and Edoardo M. Airoldi}
\begin{document}
\maketitle
\begin{abstract}

%%% The left and right page headers are defined here:
%\markboth{S. Lunag\'omez, M. Papamichalis, P.J. Wolfe, \and E.M. Airoldi}{Evaluating and Optimizing Network Sampling Designs}

%% Here are the title, author names and addresses
%\title{Evaluating and Optimizing Network Sampling Designs: Decision Theory and Information Theory Perspectives}

%\author{Sim\'on Lunag\'omez}
%\affil{Lancaster University, Lancaster LA1 4YW, United Kingdom \email{s.lunagomez@lancaster.ac.uk}}

%\author{Marios Papamichalis}
%\affil{Purdue University, West Lafayette, IN 47907, U.S.A.}

%\author{Patrick J. Wolfe}
%\affil{Purdue University, West Lafayette, IN 47907, U.S.A.}

%\author{\and Edoardo M. Airoldi}
%\affil{Temple University, Philadelphia, PA 19122, U.S.A. \email{airoldi@temple.edu}}

%\maketitle

%\begin{abstract}

Some of the most used sampling mechanisms that implicitly leverage a social network depend on tuning parameters; for instance, Respondent-Driven Sampling (RDS) is specified by the number of seeds and maximum number of referrals. We are interested in the problem of optimizing these sampling mechanisms with respect to their tuning parameters in order to optimize the inference on a population quantity, where such quantity is a function of the network and measurements taken at the nodes. This is done by formulating the problem in terms of decision theory and information theory, in turn. We discuss how the approaches discussed in this paper relate, via theoretical results, to other formalisms aimed to compare sampling designs, namely sufficiency and the Goel-DeGroot Criterion. The optimization procedure for different network sampling mechanisms is illustrated via simulations in the fashion of the ones used for Bayesian clinical trials.
\end{abstract}

%\begin{keywords}
%Decision Theory, Bayesian Experimental Design, Respondent-Driven Sampling, Social Network.
%\end{keywords}

\section{Introduction}
Respondent driven sampling (RDS) is a widely used sampling mechanism that takes advantage of social network structure. It was proposed by \cite{Hecka} and it is implemented with the aim of sampling from hidden populations, as it happens in problems from Epidemiology and Marketing. The distribution of RDS is specified by the sample size and two tuning parameters; these parameters are: the number of maximum referrals per individual and the number of starting points (also known as \emph{seeds}). If we assume that prior information is available regarding the joint distribution of social network structure and the distribution of the responses (\emph{i.e.}, observations at the node level), then it is of interest (from a methodological and a practical point of view) how such information could be used in to calibrate the RDS tuning parameters. Note that, for this question to make sense, it is necessary to specify criteria for evaluating the performance of different sampling mechanisms.\\

We argue that Decision Theory, and more specifically, Lindley's formulation of Bayesian experimental design (\cite{Lindley}), is the appropriate formalism for this task if the inference can be specified beforehand. Decision theory allows us to evaluate a design based on the average quality of the inference, here the `quality of the inference' is encoded in the loss function and the average is taken with respect to the prior predictive distribution.\\

Still, there are cases where it is not feasible to specify the inference beforehand. Therefore, a different perspective is also needed. We adopt the following rationale: a sampling mechanism should be preferred over another if it tends to have a better performance for a wide array of inferences. We make the case that casting the problem of performing Bayesian inference over a partially observed network as a data compression problem will provide: i) additional insights from a conceptual point of view and ii) robust comparisons between designs. Under this perspective, the best design is the one that preserves the most information from the process that generated the full data set.\\

In this paper, we illustrate the use of Decision Theory in calibrating the tuning parameters of sampling mechanisms on networks. As a by-product, we provide intuition about the usefulness of more general versions of RDS mechanisms. The approaches discussed in this paper enable the statistician to compare designs with the same functional form (Decision Theory) as well as designs with different functional form (Data Compression).\\

We discuss how the Decision Theory and Data Compression approaches relate to other formalisms that aim to compare sampling designs, namely sufficiency and the Goel-DeGroot criterion. We make the case that the insights provided by these perspectives can complement each other. This is done in the context of network data.

\subsection{Related work}
Our work is related to the work of \cite{ChaloVerd}, since, like them, we discuss how to implement ideas from Bayesian experimental design to solve applied problems. We adopt the formulation of Bayesian experimental design proposed by \cite{Lindley} and the framework for sequential decision making proposed by \cite{Bell}. For developing a ranking of designs based on information theory, we rely on the framework of data compression \cite{MezaMont} and the notion of a reference prior \cite{Bern}. In the context of social network data and, more specifically, the performance of RDS, our work relates to the simulation studies performed by \cite{BlitzNest}. To incorporate the different sources of uncertainty, we elaborate on the ideas developed in the technical report by  \cite{LunagAirol} and on the Markov Chain Monte Carlo approaches developed by \cite{MollPett} and \cite{AndrieuRob}.

\subsection{Contributions}
The main contributions of this paper are: First, we propose a new approach for comparing sampling designs based on the concept of data compression from information theory, thus providing a principled approach, for example, to calibrate tuning parameters of existing mechanisms and to evaluate new mechanisms; second, we discuss different aspects of framing the comparison of sampling designs on networks (which include the specification of loss functions and sequential analysis). This discussion is conducted around a fully Bayesian model for RDS, which is novel; third, we provide theoretical results that show the usefulness and limitations of sufficiency, data compression and the criterion of Goel-DeGroot for comparing sampling designs on networks; and fourth, we propose  new network sampling mechanisms from our understanding of of these new principled approaches and carry out comparative performance evaluation.

\section{Problem Set-up}
\subsection{Respondent-Driven Sampling}\label{Sec:RDS}
Let $\mathcal{G}$ be a social network with $N$ nodes and assume those nodes are labeled. Let $Y$ denote a vector that has as $i$-th component (denoted by $Y(i)$) a measurement to be taken at the $i$-th node, here $1\leq i \leq N$. The objective is to perform inference on a feature $Q$ of the joint distribution of $(\mathcal{G},Y)$ (for instance: the average probability that $\left\{Y(i)=1\right\}$ for $Y(i)$ binary with possible values in $\left\{ 0,1\right\}$, $1\leq i \leq N$). The network $\mathcal{G}$ and the responses $Y(i)$, $1\leq i \leq N$, are accessible only trough sampling. Respondent-Driven Sampling (RDS) is a procedure, proposed by \cite{Hecka}, that deals with this problem. It is defined as a set of policies that allow the sampling to propagate trough the network, conditional on a set of starting points or \emph{seeds}.\\

Respondent-Driven Sampling can be understood as a stochastic process on discrete time that is conditional on the underlying network $\mathcal{G}$ and has as its state space $(v_{[t]},\mathcal{G}_{I[t]},Y_{I[t]},\check{D}_{I[t]})$; where $v_{[t]}$ denotes the labels of nodes recruited at time $t$; $\mathcal{G}_{I[t]}$ denotes the subgraph of $\mathcal{G}$ implied by the nodes recruited up to time $t$ and the edges that encode the information about which node from time $s-1$ recruited which node at time $s$, where $1\leq s \leq t$; $Y_{I[t]}$ and $\check{D}_{I[t]}$ denote, respectively, the observed responses and the reported degrees up to time $t$. \\

The way the sampling propagates trough the network is defined by the following policies:
\begin{enumerate}
\item Sample $w_0$ nodes uniformly from $\mathcal{G}$. This is known as the $0$-\emph{th wave}, its elements are commonly known as \emph{seeds}. The selected nodes constitute $\mathcal{G}_{I[0]}$.
\item For each node $i$ in Step 1, record the response $Y(i)$ in $Y_{I[0]}$ and the corresponding reported degree $\check{D}(i)$ in $\check{D}_{I[0]}$. 
\item For each node in the $(k-1)$-\emph{th wave}, sample uniformly $m$ nodes among its neighbours relative to $\mathcal{G}$ and such that they have not been sampled before. This is known as the $k$-\emph{th wave}. The indices for these nodes constitute $v_{[k]}$.
\item For each node $i$ sampled in Step 3, record the response $Y(i)$ in $Y_{I[k]}$, the corresponding reported degree $\check{D}(i)$ in $\check{D}_{I[k]}$, and the edge that connected $i$ to $\mathcal{G}_{I[k-1]}$ to construct $\mathcal{G}_{I[k]}$.
\item Repeat Steps 3 and 4 until the pre-specified sample size $n$ has been attained. Interrupt the current wave if necessary.
\end{enumerate}
Observe that Step (a) was set this way for the sake of simplicity, since our interest is on evaluating sampling procedures. The distribution of the starting points can be modified depending on the question at hand. Clearly, the notion of wave encodes the discrete time involved in the sampling process. A design that follows these policies, except for (d) is called \emph{link tracing}.\\
 
Given the sample size $n$, RDS has two tuning parameters: $m$, which denotes the number of referrals; and $w_0$, which denotes the number of seeds. It is of interest to calibrate $\eta=(m,w_0)$ by taking into account the type of inference that will be performed and any prior assumptions on the joint distribution of $\mathcal{G}$ and $Y$. We will use the notation $I_{\text{RDS}}$ and $I_{\text{LT}}$ to refer to RDS and link tracing, respectively, when more than one sampling design is being discussed. 

\subsection{Notion of Non-Ignorability}\label{Sec:NonIgno}
\begin{definition}[Ignorability of a Sampling Design] Let $Z$ denote the full data and $Z_{\text{INC}}$ represent the observed data. Let $p(Z \mid \tau)$ denote the distribution for the full data. A sampling mechanism $I$ is \emph{ignorable} if
\begin{displaymath}
p(I \mid Z,\eta)=p(I \mid Z_{\text{INC}},\eta),
\end{displaymath}
and the parameters for the sampling mechanism ($\eta$) and the full data ($\tau$) are distinct. If a sampling mechanism is ignorable, then the term corresponding  to the distribution of $I$ can be omitted from the likelihood. 
\end{definition}

In the context of this paper $Z_{\text{INC}}$ will usually be constituted by $\mathcal{G_\text{INC}}$ and $Y_\text{INC}$, which denote, respectively, the observed part of the network and the observed responses for the nodes. We will use $\mathcal{G_\text{EXC}}$ and $Y_\text{EXC}$ to denote, respectively, the unobserved part of the network and the unobserved responses for the nodes. Here $\text{INC}$ and $\text{EXC}$ stand for `included' and `excluded' from the sample.\\

Two well-known examples of ignorable designs on networks are snowball \citep{Good} and ego-centric \citep{NewEgo}, which are denoted by $I_{\text{Snow}}$ and $I_{\text{Ego}}$, respectively. They are defined as follows:
\begin{enumerate}
\item Ego-centric design of sample size $n$ is performed by choosing $n$ elements of $\mathcal{V}$ uniformly at random; this constitutes $\mathcal{V}_{\text{INC}}$. Then, we let
 $\mathcal{G_\text{INC}}$ be defined by the edges and non-edges associated to the rows and colummns of $A_{\mathcal{G}}$ corresponding to $\mathcal{V}_{\text{INC}}$. 
\item Snow-ball design with $k$ waves and $s$ seeds is performed by: first, picking $s$ elements of $\mathcal{V}$ uniformly at random (these will be the seeds), as with RDS, this is the $0th$ wave (the seeds are added to $\mathcal{V}_{\text{INC}}$); second, include all the edges and non-edges of the $i$-th wave into $\mathcal{G_\text{INC}}$; third, let the nodes incident to the edges observed in the second step be added to $\mathcal{V}_{\text{INC}}$, this is the $(i+1)$-th wave;  fourth, we increment $i$ by one and repeat the second and third steps until the target number of waves is reached or the network is exhausted.
\end{enumerate}

In the case of RDS, the distribution of the sampling mechanism $I$ given the full data $Z=(Y,\mathcal{G})$ and the tuning parameter ($\eta$) is given by:
\begin{equation}\label{Eq:Lik4RDS}
p(I \mid \mathcal{G}, \eta)=\frac{1}{\binom{\tilde{d}_0}{m}} \left( \prod_{j_1=1}^{w_0} \frac{1}{\binom{\tilde{d}_{j_1}}{m}}  \left[     \prod_{j_2,j_1=1}^{w_{j_1}} \frac{1}{\binom{\tilde{d}_{j_2,j_1}}{m}}   \cdots   \left[    \prod_{j_k,\dots,j_1=1}^{w_{j_{k-1},\dots,j_1}}   \frac{1}{\binom{\tilde{d}_{j_k,\dots,j_1}}{m}}    \right]   \cdots  \right]   \right). 
\end{equation}
For the sake of simplicity, we present the functional form of this distribution where the number of seeds is equal to one. Here $\tilde{d}_{(\cdot)}$ denotes the number of adjacent nodes to a given vertex (with respect to $\mathcal{G}$) that have not been sampled yet; we refer to this quantity as the \emph{adjusted degree}; in particular $\tilde{d}_0$ is the adjusted degree of the seed. $w_{(\cdot)}$ denotes the number of recruited individuals by a given node during the previous wave, while $k$ represents the number of waves needed to recruit $n$ individuals; $m$ denotes the maximum number of referrals. Here $\eta=(m,w_0)$. It is assumed that each node in waves $\left\{0,1,\dots, k-1 \right\}$ will try to recruit uniformy from the set of available networks (therefore the binomial coefficients); the products are over the nodes recruited at each wave. Most of the modified versions of RDS discussed in this paper will imply a similar expression for $p(I \mid \mathcal{G},\eta)$.\\

Observe that RDS is ignorable when the vector of $\tilde{d}$'s is fully observed. There are situations, that often arise in practice, that prevent this from happening, for example:
\begin{enumerate}
\item The degrees are reported with noise, this is common in Epidemiology, more specifically in HIV studies. Populations such as men that have sex with men tend to round the number of sexual partners they had. In this context the rounding tends to be coarser as the true number of partners gets higher. This phenomenon is known as \emph{heaping}.
\item The degrees are reported exactly, but the number of neighbours in the network that have not been sampled yet (\emph{i.e.,} the adjusted degree) is unknown.
\end{enumerate}
The methodology we propose in this paper is able to calibrate $\eta$ even when $I$ is non-ignorable. This is possible since we adopt a model that takes into account the main sources of uncertainty for dealing with this issue. The functional form presented in Equation \ref{Eq:Lik4RDS} is key for the formulation of a fully probabilistic model for RDS and the extensions of that design to be discussed in Section \ref{Sec:OldAndNew}.

\subsection{A Realistic Model of Respondent-Driven Sampling}\label{Sec:LunaGAiroldi}
We assume a probabilistic model of the form:
\begin{equation}\label{Eq:LAModel}
p(\mathcal{G},Y,I, \alpha, \gamma)=p(\mathcal{G} \mid \alpha)p(\alpha)p(I \mid \mathcal{G}, \eta) p(Y\mid \mathcal{G}, \gamma)p(\gamma).
\end{equation}
Here $\mathcal{G}$ denotes the social network, which is assumed to be a realization of a random graph (statistical network) model with parameter $\alpha$. $I$ denotes the sampling mechanism, which is understood as a probabilistic process that propagates through $\mathcal{G}$ and is determined by a set of policies $\eta$ (the tuning parameters of the design). $Y$ denotes a response vector, the response $Y(i)$ is associated to the $i-$\emph{th} node of $\mathcal{G}$. The joint distribution of the $Y$ vector is assumed to be specified in terms of $\mathcal{G}$ (using a Markov random field formulation) and a parameter $\gamma$ which controls the `strength' of the dependence among units. We define $\mathcal{G}_{\text{INC}}$ as the observed portion of $\mathcal{G}$ conditional on $I$; denote by $\mathcal{G}_{\text{EXC}}$ the unobserved portion of the network. We define $Y_{\text{INC}}$ and $Y_{\text{EXC}}$ in an analogous manner. Here $p(\alpha)$ and $p(\gamma)$ are, respectively, the priors for $\alpha$ and $\gamma$.\\

%Now, in terms of the notation presented at the beginning of last section:
%\begin{displaymath}
%\theta=(\alpha,\gamma,\mathcal{G}_{EXC},Y_{EXC})
%\end{displaymath}
%and
%\begin{displaymath}
%z=(\mathcal{G}_{INC},Y_{INC},I).
%\end{displaymath}
%Now, let $Q(\theta)$ be the population mean for $Y$. If the goal of the study is the point estimation of $Q$, then $d=\hat{Q}$.

To ease the exposition, we assume specific distributions for the factors in Expression \ref{Eq:LAModel}.  We adopt well-understood models for the random graph $\mathcal{G}$: i) the Erd\"{o}s-R\'{e}nyi model (\cite{ErdosRenyi}); ii) the Stochastic Block model; iii) the latent space model. For the vector of responses $Y$, we assumed the Markov Random Field (MRF) implied by the following Boltzmann distribution:
\begin{equation}
\Pr(Y=y\mid \mathcal{G},\gamma)\propto \exp(\gamma_0 V_0+ \gamma_1 V_1),
\end{equation}
where
\begin{displaymath}
V_0=\sum_{i=1}^{N} y(i), \quad \text{and}\quad V_1=\sum_{\left\{ (i,j) \mid A(i,j)=1\right\}}y(i)y(j).
\end{displaymath}
Here $A$ denotes the adjacency matrix for $\mathcal{G}$ and $\gamma=(\gamma_0,\gamma_1)$. This implies that the conditional distribution of the response of node $i$ given the values of all the other responses, $\mathcal{G}$ and $\gamma$ is given by:
\begin{displaymath}
\Pr(Y(i)=y(i)\mid Y(-i)=y(-i),\mathcal{G},\gamma)\propto \exp\left(\gamma_0y(i)+ \gamma_1\sum_{\left\{ (i,j) \mid A(i,j)=1\right\}}y(i)y(j)\right).
\end{displaymath}
As in \cite{MollPett}, we assume a uniform prior on
\begin{displaymath}
\gamma\in \Gamma=[\min \gamma_0, \max \gamma_0]\times[0, \max \gamma_1].
\end{displaymath}

Let $D$ denote the vector that has as $i$-th component the degree of node $i$ with respect to $\mathcal{G}$. For this paper we consider only the case where degrees are reported exactly, \emph{i.e.}, $\check{D}=D$. Here  $\check{D}$ is partitioned via $I$ into $\check{D}_{\text{INC}}$ and $\check{D}_{\text{EXC}}$, which are, respectively, the reported degrees and the degrees that would have been reported if we had access to the corresponding nodes via sampling. Observe that, for the case $\check{D}=D$, $\check{D}_{\text{INC}}$ is a deterministic function of $\mathcal{G}$ and $I$, therefore it can be included as part of the data without the need of adding an extra factor to Expression \ref{Eq:LAModel}.\\

The computation of the posterior $p(Q\mid Y_{\text{INC}},\mathcal{G}_{\text{INC}}, \check{D}_{\text{INC}}, I)$ for the model given by Expression \ref{Eq:LAModel} is performed via Bayesian model averaging (BMA, see \cite{RafteMadigVolin} and \cite{Rober}, Section 7.4), \emph{i.e.}, $p(Q \mid Y_{\text{INC}}, \mathcal{G}_{\text{INC}},\check{D}_{\text{INC}},I)$ is equal to
\begin{equation}\label{Eq:BMA}
\sum_w p(\mathcal{G}_{{\text{EXC}},w}, \alpha_w\mid  \mathcal{G}_{\text{INC}},\check{D}_{\text{INC}},I) \int_{\Theta(\mathcal{G}_{{\text{EXC}},w})} p_w(Q \mid \theta_w,\varphi_w) p(\theta_w \mid Y_{\text{INC}}, \mathcal{G}_{\text{INC}},\mathcal{G}_{{\text{EXC}},w}) d\theta_w.
\end{equation}
Here $\theta_w=(\gamma_w,Y_{{\text{EXC}},w})$, \emph{i.e.}, the parameters of the dependence structure and the missing response data; let $\varphi_w=(\alpha_w,\mathcal{G}_{{\text{EXC}},w})$. The reason we adopted this strategy for computing the posterior is the following: Since RDS is non-ignorable, it is necessary to impute missing data in order to compute the likelihood.  Typically, the number of nodes and edges of the unobserved part of the network $\mathcal{G}_{\text{EXC}}$ is unknown, which turns this problem into one of variable dimension. BMA allows to decompose this problem into stages. The mixing distribution of the BMA $p(\mathcal{G}_{{\text{EXC}},w}, \alpha_w\mid \mathcal{G}_{\text{INC}},\check{D}_{\text{INC}},I)$ is used to determine the nodes and edges to augment to $ \mathcal{G}_{\text{INC}}$. Conditioning on the imputation for the unobserved part of the network, the problem becomes one of fixed dimension and standard MCMC techniques can be used (in particular, to deal with the updates for the MRF parameters, we used the approach proposed by \cite{MollPett}). We consider two possible choices for $p_w(Q \mid \theta_w,\varphi_w)$: the first one corresponds to the problem of estimation,
\begin{displaymath}
p_w(Q^{(i)} \mid \alpha^{(i)},\gamma^{(i)}),
\end{displaymath}
where $Q^{(i)}$ is the mean of $Y_{\text{AUG}}$, a vector of responses simulated from the predictive distribution implied by Expression \ref{Eq:LAModel};
the second one corresponds to the problem of prediction,
\begin{displaymath}
p_w(Q^{(i)} \mid Y_{{\text{EXC}},w}^{(i)}, Y_{\text{INC}}^{(i)}),
\end{displaymath}
where $Q^{(i)}$ is the mean of the vector $(Y_{{\text{EXC}},w}^{(i)}, Y_{\text{INC}}^{(i)})$. Estimation and prediction will be discussed in more detail once the associated loss functions have been introduced.
%For the quadratic loss, $\hat{Q}$ is given by the mean of the $Q^{(i)}$'s, while for the multilinear loss (Expression \ref{Eq:Multilinear}) it is given by the $(k_2/(k_1+k_2))$ fractile.

\section{Evaluating and Optimizing Network Sampling Designs: Decision Theory and Information Theory Perspectives}

\subsection{Old and New RDS-based Designs}\label{Sec:OldAndNew}
In order to calibrate the vector of tuning parameters for RDS, which is denoted by $\eta=(m,w_0)$, it is necessary to establish criteria for evaluating and comparing the sampled designs implied by different specifications for $\eta$. The main objective of this paper is to provide priciple-based tools for comparing sampling designs on networks. To motivate the discussion in the remaining sections, we discuss some examples. All of the examples described in this section can be understood as settings where there is a finite family of designs $\mathcal{H}=\left\{ \eta_1, \eta_2,\dots,\eta_l  \right\}$ and where it is of interest to find an optimal $\eta^{\star}\in \mathcal{H}$ according to pre-specified criteria that take into account the type of inference to be performed and prior information on the parameters for the probabilistic model for $(\mathcal{G},Y)$.

%\subsection{Examples}
\begin{example}\label{Ex:NCoupons}
Let RDS be the sampling mechanism that propagates through the network. As in Section \ref{Sec:RDS}, denote by $w_0$ the number of seeds and regard this quantity as specified. A relevant question in this context is how to calibrate $m$, the number of referrals for a fixed sample size $n$. Let $\mathcal{H}$ be the set of possible choices for the number of referrals $\left\{ 1,2,\dots, m_{\max}\right\}$. 
\end{example}

\begin{example}\label{EX:Seeds}
Let RDS be the sampling mechanism. As in Section \ref{Sec:RDS}, denote by $w_0$ the number of seeds and let $m$ be the number of referrals.    Consider the problem of calibrating $w_0$ for $m$ fixed and a pre-specified  sample size ,\emph{i.e.}, let $\mathcal{H}$ be the set of possible choices for the number of seeds $\left\{ 1,2,\dots, w_{0,\max}\right\}$. 
\end{example}

\begin{example}\label{Ex:Curve}
Let us consider generalisations of the RDS setting. One could question the requirement of making $m$ constant across waves. 
Let  $\mathcal{H}$ be the set of policies that determine the number of referrals. More precisely, let the family of sampling mechanisms be defined by
\begin{displaymath}
f_{\eta}(x)=\Lambda \left[ C^{\bullet} \left( \frac{1}{W^{\bullet}} x \right)^{\eta} \right], 
\end{displaymath}
where $\eta$ is an element of a finite grid $\mathcal{H} \subset \mathbb{R}^{+}$, $C^{\bullet}$ is the maximum permissible value for $m$,  $W^{\bullet}$ is a cap for the number of waves, and 
\begin{displaymath}
\Lambda(z)=\min \left\{k\in \mathbb{N} : z\leq k \right\}.
\end{displaymath}
It is assumed that $x \in \left\{ 1,2,\dots, W^{\bullet}  \right\}$ and the sample size $n$ is fixed. 
\end{example}

\begin{example}\label{EX:Switch}
Let us consider a different generalisation of RDS. Think of a design were $m$ could only take two values: $\lambda_L$ and $\lambda_H$, where $\lambda_L < \lambda_H$. 
As in Example \ref{Ex:Curve}, one could adopt the convention that $m$ as a function of wave is non-decreasing.  
It seems reasonable to set the sample size $n$ as fixed and $W^{\bullet}$ as a dependent variable. Let $\mathcal{H}=\left\{\eta_1,\eta_2,\dots,\eta_{k_{\max}}   \right\}$ be the set of possible sampling mechanisms, more precisely: If the design $\eta_k$ is adopted, then $m=\lambda_L$ for the first $k-1$ waves and $m=\lambda_H$ for the following $W^{\bullet}-k+1$ waves.
\end{example}

\begin{example}\label{EX:Secuential}
For RDS, the number of seeds $w_0$ and the maximal number of referrals $m$ can be updated sequentially: First, $w_0$ is specified, based on prior knowledge. The second step consists in optimising $m$ based on the data collected on the $0$-\emph{th} wave; this is done while keeping the sample size $n$ fixed. Here $\mathcal{H}=\left\{(m,w_0)  \mid  1\leq m \leq m_{\max} , 1 \leq w_0 \leq w_{0,\max}\right\}$.
\end{example}

\subsection{Decision Theoretic Formulation of the Optimal Design Problem}\label{SubSec:Decision}
The framework of Bayesian experimental design (see \cite{Lindley} and the review by \cite{ChaloVerd}) allows the statistician to phrase the problem of specifying features of the experiment (assuming that they  are under the control of the practitioner) in terms of Decision Theory. For problems involving social networks, tuning the parameters of the sampling mechanism is a key aspect of the design; that is where the focus of our discussion will be. We first introduce some notation: 
\begin{enumerate}
  \item Let $\theta$ denote the parameters of the statistical model; $\theta$ is an element of a parameter space $\Theta$.
  \item We denote by $z$ a potential data set, which belongs to a sample space $\mathcal{Z}$.
  \item Let $a$ denote the decision or inference, which belongs to an action space $\mathcal{A}$.
  \item Let $\eta$ represent a specific sampling design, which belongs to a family of designs $\mathcal{H}$.
  \item The letter $\mathcal{L}$ represents different loss functions. 
   \end{enumerate}
The loss function $\mathcal{L}$ is the component of this formalism that quantifies the quality of an inference $a \in \mathcal{A}$, given $\theta \in \Theta$ and data $z \in \mathcal{Z}$. It is required for $\mathcal{L} $ to be non-negative, and such that it takes the value zero when the inference is correct (\emph{e.g.}, when the estimated value of $\theta$ is equal to $\theta$). Remember that we are assuming a sampling mechanism of the form $p(I \mid Z,\eta)$. As in Section \ref{Sec:LunaGAiroldi},  $I$ partitions the full data $z\in\mathcal{Z}$ into $z_{\text{INC}}$ and $z_{\text{EXC}}$, which denote, respectively, the observed and unobserved parts of the data. Let $\mathcal{Z}_{\text{INC}}$ be the space of potential $z_{\text{INC}}$'s allowed by a given sampling design. From a decision theoretic point of view, Bayesian inference is conditional on the data $z_{\text{INC}} \in \mathcal{Z}_{\text{INC}}$ and it is given by the argument in the action space $a \in \mathcal{A}$ that minimises 
   \begin{equation}\label{Eq:ExpLoss}
\mathbb{E}\left( \mathcal{L}\left(a\left(z_{\text{INC}}\right),Q(\theta)\right) \mid z_{\text{INC}} \right)=\int_{\Theta}  \mathcal{L}(a(z_{\text{INC}}),Q(\theta))p(\theta \mid z_{\text{INC}})  d\theta.
     \end{equation}
We adopted this notation to emphasise that the inference $a$ is a function of the data $z_{\text{INC}}$ and that the object of the inference $Q$ is a function of either parameters or missing data. According to the formulation proposed by  \cite{Lindley}, the loss associated  to a design $\eta \in \mathcal{H}$ is given by the average expected loss of the optimal inference over all possible data sets. This is:
   \begin{equation}\label{Eq:ExpLossDes}
   \mathcal{L}(\eta)= \int_{\mathcal{Z}_{\text{INC}}} \min_{a \in \mathcal{A}} \int_{\Theta}  \mathcal{L}(a(z_{\text{INC}}),Q(\theta))p(\theta \mid z_{\text{INC}})  d\theta  p(z_{\text{INC}}\mid \eta)dz_{\text{INC}}.
   \end{equation}
   Therefore, the optimal design is defined as the element in $\mathcal{H}$ that minimises Expression \ref{Eq:ExpLossDes}:
    \begin{equation}\label{Eq:OptiDes}
   \mathcal{L}(\eta^\star)=\min_{\eta \in \mathcal{H}} \int_{\mathcal{Z}_{\text{INC}}} \min_{a \in \mathcal{A}} \int_{\Theta}  \mathcal{L}(a(z_{\text{INC}}),Q(\theta))p(\theta \mid z_{\text{INC}}) d\theta  p(z_{\text{INC}}\mid \eta)dz_{\text{INC}}.
   \end{equation}
   In the context given by the model discussed in Section \ref{Sec:LunaGAiroldi}:
   \begin{equation}
   z_{\text{INC}}=(Y_{\text{INC}},\mathcal{G}_{\text{INC}},\check{D}_{\text{INC}},I)\quad \text{and} \quad \theta=(\alpha,\gamma,Y_{\text{EXC}},\mathcal{G}_{\text{EXC}}).      
   \end{equation}
   Therefore, $ p(\theta \mid z_{\text{INC}})$ is the joint posterior for the model parameters and missing data implied by Expression \ref{Eq:LAModel}. To compute Expression \ref{Eq:ExpLoss}, only a slight modification of Expression \ref{Eq:BMA} is needed, more precisely:
   \begin{equation}\label{Eq:ExpLossLunaGAiroldi}
 \mathbb{E}\left( \mathcal{L}\left(a\left(z_{\text{INC}}\right),Q(\theta)\right) \mid z_{\text{INC}} \right)=\sum_w p(\mathcal{M}_w\mid  z_{\text{INC}}) \int_{\Theta(\mathcal{M}_w)}   \mathcal{L}(a(z_{\text{INC}}),Q(\theta_w)) p(\theta_w \mid z_{\text{INC}}) d\theta_w.
\end{equation}
   Within this context, $ p(z_{\text{INC}}\mid \eta)$ is the prior predictive distribution of $(Y_{\text{INC}},\mathcal{G}_{\text{INC}},\check{D}_{\text{INC}},I)$ implied by the model given in Expression \ref{Eq:LAModel}. It is computed by performing the following steps:
   \begin{enumerate}
   \item Generate a sample from the distribution $p(\mathcal{G}, Y, \alpha,\gamma)=p(\mathcal{G} \mid \alpha)p(\alpha)p(Y\mid \mathcal{G},\gamma)p(\gamma)$.
   \item Given $\mathcal{G}$, simulate $I \mid ( \mathcal{G},\eta )$ and let 
\begin{displaymath}
z_{\text{INC}}=(Y_{\text{INC}},\mathcal{G}_{\text{INC}},\check{D}_{\text{INC}},I).
\end{displaymath}
   \end{enumerate} 
   
   We will focus on the case where the inference of interest is estimation. This implies $Q(\theta)=Q(\alpha,\gamma)$. For this problem two loss functions are particularly relevant: the quadratic loss
   \begin{displaymath}
    \mathcal{L}(a(z_{\text{INC}}),Q(\theta))=(a(z_{\text{INC}})-Q(\theta))^2,
   \end{displaymath}
   for which the optimal decision is given by the posterior mean, and the multilinear loss
   \begin{equation}\label{Eq:Multilinear}
   \mathcal{L}(a(z_{\text{INC}}),Q(\theta))=
   \begin{cases}
    k_2(Q(\theta)-a(z_{\text{INC}}))          & \text{ if } \theta>a,\\
    k_1(a(z_{\text{INC}})-Q(\theta))          & \text{ otherwise. }
   \end{cases}
   \end{equation}
   For this loss, the optimal decision is the $(k_2/(k_1+k_2))$ fractile of the posterior (see Section 2.5 of \cite{Rober}). \\
   
   Sometimes the object of the inference is prediction. In this paper we discuss two approaches for dealing with this problem from the Decision Theory perspective: the first approach is to adopt a loss function that encodes the gain of information for the predictive distribution of the quantity of interest $Q(z)$ due to sampling. Such gain of information is measured via the Kullback-Leibler divergence between the prior $p(Q(z))$ and posterior $p(Q(z_{\text{AUG}},z_{\text{INC}})\mid z_{\text{INC}})$ predictive distributions of $Q(z)$, \emph{i.e.}, 
   \begin{displaymath}
    \mathcal{L}(\eta)=\sum_{z_{\text{INC}} \in \mathcal{Z}_{\text{INC}}}-\sum_q \log\left(\frac{p(Q(z_{\text{AUG}},z_{\text{INC}})=q\mid z_{\text{INC}})}{p(Q(z)=q)}\right)p(Q(z_{\text{AUG}},z_{\text{INC}})=q\mid z_{\text{INC}}) p(z_{\text{INC}}\mid \eta),
   \end{displaymath}
   where $z_{\text{AUG}}$ denotes an imputed value for $z_{\text{EXC}}$ obtained from the posterior predictive distribution. The second approach for evaluating a design in terms of prediction is to use what is called an \emph{intrinsic loss}, which measures the distance between the predictive distribution of the quantity of interest $Q(z)$ given the parameter $\theta$ with respect to the distribution implied by the inference $a(z_{\text{INC}})$. For the sake of concreteness we will use the Hellinger distance:
   \begin{displaymath} 
    \mathcal{L}(a(z_{\text{INC}}),Q(\theta))=\frac{1}{2}\sum_q \left( \sqrt{\frac{p(Q(z_{\text{AUG}},z_{\text{INC}})=q\mid z_{\text{INC}},\theta)}{p(Q(z_{\text{AUG}},z_{\text{INC}})=q\mid z_{\text{INC}},a(z_{\text{INC}}))}}  -1 \right)^2   p(Q(z_{\text{AUG}},z_{\text{INC}})=q\mid z_{\text{INC}},\theta).
    \end{displaymath}
% \section{Model of Lunag\'omez-Airoldi}  

%\begin{enumerate}
%\item the optimal design $\eta^\star$ based on prior information regarding the parameters and the type of inference to be performed once the data set is collected.
%\item the optimal inference $a^\star$ based on the posterior implied by the data $z_{\text{INC}}$ and the loss function $\mathcal{L}$. The data is $z_{\text{INC}}$ is obtained using the sampling mechanism entailed by $\eta^\star$.
%\end{enumerate}

\subsection{Comparing Sampling Designs via Data Compression}\label{Sec:DesignCompress}
Given a set of potential sampling designs 
$\left\{ I_1,I_2,\dots, I_{\text{k}}  \right\}$ to be applied to an observation $\mathcal{G}$ from a random graph model $p(\mathcal{G}\mid \alpha)$, the objective is to produce a ranking of these designs such that higher positions in the ranking lead to posteriors 
$p(\alpha \mid\mathcal{G}_{\text{INC}})$ that preserve more information about the probabilistic mechanism that generated the data. To achieve this, we cast the problem of performing Bayesian inference from a partially observed graph as an instance of data compression. The optimal design is the one that minimizes the loss of information.  \\

The problem of data compression can be summarized as follows: An observation $x$ from a random variable $X$ defined on $\mathcal{X}$ is obtained. The relevant notion of size for an observation is the length of a string with entries in $\left\{ 0,1\right\}$ needed to represent it. To transmit $x$, we transform it into $y_x\in \mathcal{Y}$, where $y_x$ has smaller size than $x$; this process is called \emph{data compression}. To retrieve as much as possible of $x$, $y_x$ is transformed to $\tilde{x}\in\mathcal{X}$; this process is called \emph{data decompression}\\

To illustrate the main ideas, we assume that $\alpha$ is fixed but unknown, with value $\alpha^*$. Let $\mathcal{G}$ be an observation from $p(\mathcal{G}\mid \alpha^*)$. The act of applying $I$ to $\mathcal{G}$ corresponds to compressing a random adjacency matrix
into  $\mathcal{G}_{\text{INC}}$. From this partially observed network,  a posterior distribution $p(\alpha\mid \mathcal{G}_{\text{INC}})$ is computed. The act of computing the posterior and the posterior predictive $p(\mathcal{G}\mid \mathcal{G}_{\text{INC}})$  corresponds to decompressing the adjacency matrix.
To evaluate the loss of information, we compute the Hellinger distance between the distribution of the full data entailed by $\alpha^*$  and the posterior predictive implied by $\mathcal{G}_{\text{INC}}$, \emph{i.e.},
\begin{displaymath}
\text{Loss of information}=H(p(\mathcal{G} \mid \alpha^*), p(\mathcal{G} \mid \mathcal{G}_{\text{INC}})).
\end{displaymath}
By integrating with respect to $p(\mathcal{G}\mid \alpha)$, we obtain the average of those distances: 
\begin{equation}
\psi_\bigstar (I,\alpha^*)=\sum_{\mathcal{G}_{\text{INC}}} \sum_{\mathcal{G}_{\text{EXC}} \sim \mathcal{G}_{\text{INC}}}   H(p(\mathcal{G}_{\text{INC}},\mathcal{G}_{\text{EXC}} \mid \alpha^*), p(\mathcal{G}_{\text{INC}},\mathcal{G}_{\text{EXC}} \mid \mathcal{G}_{\text{INC}})),
\end{equation}
where $\mathcal{G}_{\text{EXC}} \sim \mathcal{G}_{\text{INC}}$ denotes the statement that $\mathcal{G}_{\text{EXC}}$ provides the information needed for $ \mathcal{G}_{\text{INC}}$ to become an adjacency matrix. This statement is context dependent (\emph{e.g.}, if there a maximal number of nodes $N$ to consider, or if $ \mathcal{G}_{\text{INC}}$ contains information about ). The different values for $\mathcal{G}_{\text{EXC}}$ will be weighted differently for $p(\mathcal{G}_{\text{INC}},\cdot \mid \alpha^*)$ and $p(\mathcal{G}_{\text{INC}},\cdot \mid \mathcal{G}_{\text{INC}})$ . $\psi_\bigstar (I,\cdot)$ will be the score associated to the amount of information preserved by the design $I$.\\

%insight from reference prior
%Insight from entropy
$ \mathcal{G}_{\text{INC}}$
 We incorporate uncertainty on $\alpha$ by assuming it to be a sample from a distribution $f(\cdot\mid \alpha^*)$ with mean $\alpha^*$. If this distribution coincides with the prior, the setting becomes a particular case of the one discussed in Section \ref{SubSec:Decision}. However, by making $f(\cdot\mid \alpha^*)$ different from $p(\alpha)$, we can get the following insights:
 \begin{enumerate}
     \item While specifying $\alpha$ leads to the computation of a risk function and assuming a prior leads to the computation of the expected loss, it is useful to investigate the performance of a design for regimes (regions, intervals) of the parameter space.
     \item It provides an understanding of how the Bayesian experimental design behaves when the prior is misspecified.
 \end{enumerate}
The score for $I$ becomes:
\begin{equation}\label{Eq:InfoScore}
\psi(I,f(\cdot\mid \alpha^*))=\int \psi_\bigstar (I,\alpha) f(\alpha\mid \alpha^*)\text{d}\alpha
\end{equation}

One question that arises when sampling from a network can be phrased as follows: How many entries of the adjacency matrix have to be observed so that the inference from that sample approximates the inference obtained from a full graph? To get a better understanding of this, we focus on a similar question: To what extent one can reconstruct a full network based on a compressed version of it?  The following result, as stated in \cite{MezaMont}, provides insight regarding how many entries of an adjacecy matrix need to be observed in order to reconstruct a graph $\mathcal{G}$ sampled from a model $p(\cdot\mid \alpha)$.

\begin{thm}Let $L^*$ denote the shortest average length that can be achieved by compressing a random vector $X$ via a code. Let $H_{X}$ denote the entropy of a random vector $X$. Then, for any $n\in \mathbb{N}^+$
\begin{displaymath}
H_X\leq L^*\leq H_X+1
\end{displaymath}\label{thm:Compress}
\end{thm}

We now present two particular instances of this result:
\begin{prop}\label{Prop:Entropy}
The entropy for Erd\"{o}s-R\'{e}nyi with parameters $(N,\alpha)$ is given by
\begin{displaymath}
H_{\text{ER}}=-\frac{N(N-1)}{2}\left[ \alpha \log(\alpha) + (1-\alpha) \log(1-\alpha)\right]
\end{displaymath}
in addition, the entropy for the stochastic block model with $K$ blocks is given by:
\begin{displaymath}
H_{\text{SBM}}=2\times N \times H_{\text{Block}}+\binom{N}{2}\times\tilde{H}_{\text{Inclusion}},\nonumber
\end{displaymath}
where
\begin{eqnarray}
H_{\text{Block}}& = & -\sum_{i=1}^K \beta_i \log(\beta_i)\nonumber\\
\tilde{H}_{\text{Inclusion}} & = & -\sum_{\left\{i,j\right\}\in \left\{1,2,\dots,K \right\}} \beta_i\beta_j\left[  \alpha_{i,j}\log(\alpha_{i,j})+(1-\alpha_{i,j})\log(1-\alpha_{i,j}) \right],\nonumber.
\end{eqnarray}
here, $\beta_i$ denotes the probability of allocating a node to block $i\in\left\{ 1,\dots,K\right\}$ and $\alpha_{i,j}$ the probability of including an edge between a node allocated to block $i$ and a node allocated to node $j$.
\end{prop}
The proof is provided in Appendix 1. It is worth mentioning that this results are meant to provide lower bounds for sample sizes. Theorem \ref{thm:Compress} was meant for compression of the whole adjacency matrix; sampling designs on networks are a constrained version of this.

%where $\tau(\cdot)$ denotes the operation of extracting a specific feature from a network (\emph{e.g.} degree distribution, modularity) and $p(\tau(\mathcal{G}))$ is the predictive distribution for the feature implied by this prior distributions.

\subsection{Optimal Bayesian Experimental Design via Dynamic Programming}
Lindley's Formulation is a particular case of two-stage finite decision problem (\cite{Bell}, Chapter 3 and \cite{ParmiInoue}, Section 12.3) since two decisions are taken sequentially: First, a decision regarding the design $\eta$ before any data is observed. After the data $z_{\text{INC}}$ has been observed, a decision regarding the specific inference $a(z_{\text{INC}})$ is performed. In this context, it is clear that the first decision imposes constraints on the potential data sets to be observed in the future, and also affects the value of the final Bayesian inference.  
Dynamic programming allows to generalise Lindley's Formulation for the case where multiple decisions can be taken during the data collection process, by this it is meant that the practitioner will alternate between observing data and making a decision regarding the design, where such decision will be based on the posterior obtained from the data observed up to that point in time.
\\

\begin{figure}[!ht]
\begin{center}
  \includegraphics[width=0.85\textwidth]{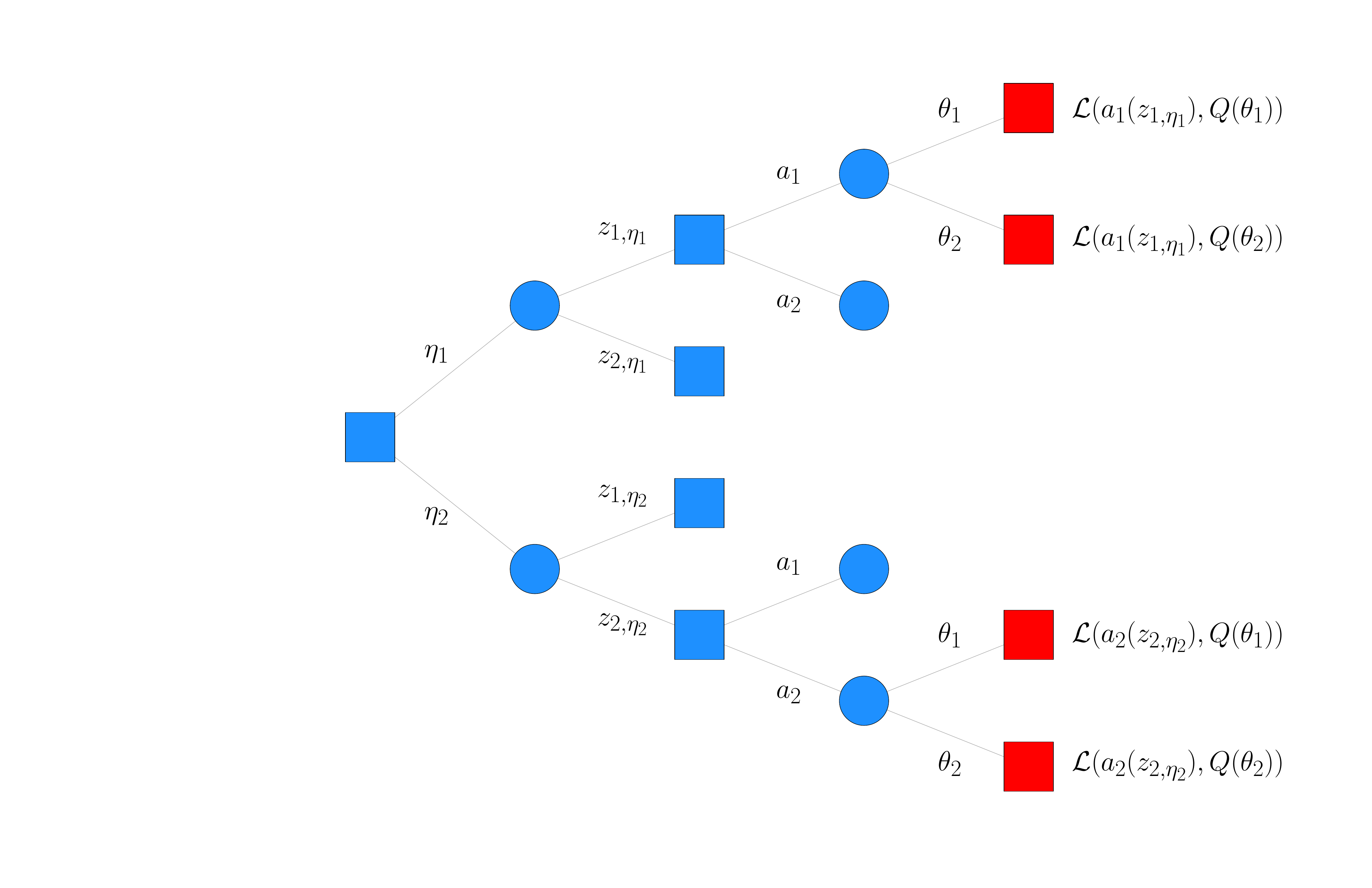}
  \caption{Lindley's Formulation illustrated as a general two-stage decision tree. }\label{Fig:DecisionTree}
  %The tree is rooted at the node associated to the decision of assigning a specific value to $\eta$.  Each leaf of this tree is associated to the value of the loss function corresponding to the path from the root to the leaf. }\label{Fig:DecisionTree}
\end{center}
\end{figure}

The two-stage decision problem is often visualized as a layered tree. The layers encode the temporal sequence involving decisions and data collection, more precisely, if two events (\emph{i.e.}, the nodes of the tree) are connected by an edge, the one located at the layer on the left precedes the one located at the layer on the right. Decisions are usually represented by squares and data collection events (which include the final hypothetical disclosure of the true value of the parameters of the model ) are represented by circles. The labels for the edges of the tree indicate the possible decisions and data sets involved in the decision problem. In this paper we adopt all of these conventions. An example of such decision tree is displayed in Figure \ref{Fig:DecisionTree}. Figure \ref{Fig:DecisionTree} also serves to illustrate the fact that Lindley's Formulation can be understood as a two-stage finite decision problem. The event in Lindley's Formulation with highest precedence is the decision regarding the tuning parameters of the design $\eta$, this event is followed by the collection of a potential data set $z_{\text{INC}}$, the next event in this process is the inference $a(z_{\text{INC}})$, and the last step is given by the hypothetical disclosure of the quantity of interest $Q(\theta)$. Once all the decisions and information are available, the loss corresponding to each leaf of the tree can be computed; such loss is given by  $ \mathcal{L}(a(z_{\text{INC}}),Q(\theta))$.\\

The algorithm that solves the the two-stage decision problem is called \emph{backwards induction}; it was proposed by \cite{Bell}, Chapter 3. To outline the backwards induction is to rephrase the procedure proposed by Lindley (Expressions \ref{Eq:ExpLoss} - \ref{Eq:OptiDes}). The two-stage decision problem can be easily generalised (at least conceptually) to multi-stage decision problems. The backwards induction algorithm for multi-stage decision problems is described in the appendix.

\section{Theoretical Results}
While decision theory provides a principled approach for comparing sampling designs, its applicability in some situations may be limited, mainly because: i) that approach is often tailored to a specific inference, while the data may be used for different types of inferences over time; ii) it may entail an expensive amount of computing time. In this section, we explore the relationship between decision theory and other perspectives aimed to formalise the comparison between sampling designs. We do this in the context of sampling designs on network data.\\

\begin{definition}[Sufficiency Criterion] Let $I_X=\{X,S_X;P_{\theta}, \theta \in \Theta \}$ denote a statistical experiment in which a random variable or random vector $X$ defined on some sample space $S_X$ is to be observed, and the distribution $P_{\theta}$ of $X$ depends on a parameter $\theta$ whose value is unknown and lies in some parameter space $\Theta$. The experiment $I_x$ is sufficient for the experiment $I_Y$ (denoted $I_X \succeq I_Y$) if there exists a stochastic tranformation of $X$ to a random variable $Z(X)$ such that, for each $\theta \in \Theta$, the random variables $Z(X)$ and $Y$ have identical distributions. We assume that there exist generalized probability density function $p(x \mid \theta)$ for the distribution $P_{\theta}$, with respect to some $\sigma$-finite measure $\mu$.
\end{definition}

The relationship between sufficiency and decision theory is explored in \cite{Blackw} and \cite{Boll}, where the authors prove that sufficiency $I_X \succeq I_Y$ is equivalent to the situation  where $I_Y$ dominates $I_X$ in terms of Bayes risk, for all decision problems $d(\cdot,\cdot)$ and all priors on $\theta$. When used as a criterion for comparing designs, this rationale is known as the Bohnenblust, Shapley and Sherman's method.  \\

We now investigate how some sampling designs on networks compare by applying the definition of sufficiency directly. $I_{\text{Ego}}$ and $I_{\text{Snow}}$ have been defined in Section \ref{Sec:NonIgno}. Here $I_{\text{LT}}$ and $I_{\text{RDS}}$ denote, respectively, the link tracing and RDS sampling designs, as defined in Section \ref{Sec:RDS}.\\

\begin{thm}\label{Th:DesignSuff1} For number of seeds $s$, number of referrals $r$ and $w$ number of waves, all intergers $\geq 0$, for link tracing designs and respondent driven sampling designs the following hold:

\begin{enumerate}
\item $I_{\text{LT}(s,r,w+1)}\succeq I_{\text{LT}(s,r,w)}$
\item $I_{\text{LT}(s+1,r,w)}\succeq I_{\text{LT}(s,r,w)}$
\item $I_{\text{LT}(s,r+1,w)}\succeq I_{\text{LT}(s,r,w)}$
\item $I_{\text{RDS}(s,r,w+1)}\succeq I_{\text{RDS}(s,r,w)}$
\item $I_{\text{RDS}(s+1,r,w)}\succeq I_{\text{RDS}(s,r,w)}$
\item $I_{\text{RDS}(s,r+1,w)}\succeq I_{\text{RDS}(s,r,w)}$
\end{enumerate}

\end{thm}

\begin{prop}\label{Prop:DesignSuff0} For number of seeds $s$, number of referrals $r$ and $w$ number of waves and $a$, $b$, $c$, all intergers $\geq 0$, for link tracing designs and respondent driven sampling designs the following hold:

\begin{enumerate}
\item $I_{\text{LT}(s+a,r+b,w+c)}\succeq I_{\text{LT}(s,r,w)}$
\item $I_{\text{RDS}(s+a,r+b,w+c)}\succeq I_{\text{RDS}(s,r,w)}$
\end{enumerate}

\end{prop}

The proofs of these results (Appendix 1) illustrate one of the main limitations of using the notion of sufficiency to compare sampling designs on networks. To stablish if such a relationship holds between two designs, we need to find a stochastic transformation that maps the observed statistic from one design into a statistic with the same distribution as the one produced by the other design. Finding such a transformation is not always straightforward (as it happens in the proof of Theorem \ref{Th:DesignSuff1}). The alternative case is more problematic, since it entails proving that the mapping between statistics does not exist. A weaker statement would be to consider a pre-specified family of mappings only. Still, there is no canonical family of mappings between graphs to consider.  \\

\begin{thm}\label{Th:DesignSuff2} Given two designs $I_{\text{LT}(s,r,w)}$ and $I_{\text{LT}(s+1,r-1,w)}$, there is no composition of the stochastic transformations used in the proof of Theorem \ref{Th:DesignSuff1} that would stablish that one of these designs is sufficient with respect to the other.
\end{thm}

%\begin{proof}

%To construct the counter-example, we assume that the underlying graph is regular and large enough so it does not exhaust the designs.\\

%Mapping (3) decreases the expected degree of the observed subgraph, while the mapping used in (2) does not decrease nor increase the expected degree and the mapping (1) may decrease the %expected degree. Combined, these statements imply that a composition of these mappings can only decrease the expected degree.\\

%Mapping (2) decreases the number of connected components, while (1) and (3) do not affect the number of connected components. Combined, these statements imply that a composition of these mappings can only decrease the number of connected components.\\

%Therefore, a relationship of sufficiency cannot be stablished between $I_{\text{LT}(s,r,w)}$ and $I_{\text{LT}(s+1,r-1,w)}$, since $I_{\text{LT}(s+1,r-1,w)}$ will have more connected components and lower expected degree when compared to $I_{\text{LT}(s,r,w)}$.\\

%\end{proof}

Theorem \ref{Th:DesignSuff2} illustrates one of the main limitations of sufficiency as a tool for comparing sampling designs, namely that it is not always possible to establish preference between a pair of sampling designs. In contrast, the criterion proposed in Section \ref{Sec:DesignCompress} does not suffer from this drawback. We now discuss how the latter approach compares to the criterion proposed in \cite{GoelDeGr}, which also involves a KL-divergence in its formulation.

\begin{definition}[Goel-DeGroot Criterion]\label{Def:DeGroot}  Let $\Xi$ denote the class of all prior distributions on the parameter space $\Theta$. Given two prior distributions $\xi_1, \xi_2 \in \Xi$, let $p_i(x)$ denote the marginal generalized probability density function $\int_{\Theta} p(x \mid \theta) d \xi_i(\theta)$, for $i=$1,2 and let $J_X(\xi_1,\xi_2)$ denote the KL-information contained in $I_X$ for dicriminating between $p_1(x)$ and $p_2(x)$, defined by:

\begin{equation}
J_X(\xi_1,\xi_2)=\int_{S_X}p_1(x)\log \frac{p_1(x)}{p_2(x)}d\mu(x)  \nonumber
\end{equation}

If $\xi_1$ assigns probability 1 to the point $\theta=\theta_0$, we shall denote $J_X(\xi_1,\xi_2)$ by $J_X(\theta_0,\xi_2)$. The KL-information $J_Y(\xi_1,\xi_2)$ contained in $I_Y$ is defined analogously. If $I_X\succeq I_Y$ then:

\begin{equation}
J_X(\theta,\xi) \geq J_Y(\theta,\xi)\nonumber
\end{equation}

for all $\theta \in \Theta$ and $\xi \in \Xi$. Moreover:

\begin{equation}
J_X(\xi_1,\xi_2) \geq J_Y(\xi_1,\xi_2)\nonumber
\end{equation}

for all $\xi_1, \xi_2 \in \Xi$.
\end{definition}

We need to rule out the possibility that the criteria presented in Definition \ref{Def:DeGroot}  and Section \ref{Sec:DesignCompress} are equivalent. One way to prove this is to establish that the functions optimised for these criteria are not related by a monotone transformation.\\

\begin{thm}\label{Th:Monotone} The notions for comparing sampling designs proposed in Section \ref{Sec:DesignCompress} and Definition \ref{Def:DeGroot} are distinct. By this it is meant that there is no monotone transformation from one to the other.
\end{thm}

%OK, that was theory. What can we learn via concrete examples. Are there insights that we can get from one of the notions but not the other?\\

While it is interesting from a theoretical perspective that both criteria are distinct, what is relevant from the practitioner's perspective is the existence of cases, among the designs we have discussed previously, where both criteria  provide different insights. \\

\begin{prop} \label{Prop:ComprVsDeGroot} The criterion based on information theory presented in Definition \ref{Def:DeGroot} is such that:
\begin{enumerate}
\item For $I_{\text{Snow}}$ and $I_{\text{Ego}}$, it does not rank one above the other.
\item For $I_{\text{Ego}}(n_1)$ and $I_{\text{Ego}}(n_2)$ with $n_1<n_2$, it does not rank one above the other.
\end{enumerate}
\end{prop}

 The first part of the result is not surprising: since both $I_{\text{Snow}}$ and $I_{\text{Ego}}$ are ignorable, the capacity of both designs to discriminate between posterior predictive distributions should not be different. An intuitive way to see this is the property of ignorable designs that implies that once we know the values of the observed data, there is no additional information provided by the design. The issue we mean to point out arises on the computations that lead to the result: a key part of the proof of Proposition \ref{Prop:ComprVsDeGroot} is that the terms that contains information about the design cancel out. The same is true for the proof of the second part: this means that, according to the Goel-DeGroot Criterion, two ignorable sampling designs with the same functional form and different sample size cannot be ranked so one is strictly above the other. In contrast, the notion proposed in Section \ref{Sec:DesignCompress} does not suffer from this drawback.

\section{Simulation Study}
\subsection{Design of Simulation Study}
As a first step, we outline how Lindley's Formulation (\cite{Lindley} and \cite{ChaloVerd}) can be implemented via Monte Carlo. Consider the model described in Section \ref{Sec:LunaGAiroldi} and  a finite family of designs $\mathcal{H}=\left\{ \eta_1, \eta_2,\dots,\eta_l  \right\}$, then:
\begin{enumerate}
\item Generate $K$ samples from the distribution $p(\mathcal{G}, Y, \alpha,\gamma)=p(\mathcal{G} \mid \alpha)p(\alpha)p(Y\mid \mathcal{G},\gamma)p(\gamma)$ and let $\left( \mathcal{G}^{(k)}, Y^{(k)}\right)$ be the realisation of $(\mathcal{G},Y)$ associated to $k \in \left\{1,2,\dots,K \right\}$.
\item Given $\mathcal{G}^{(k)}$, simulate $I^{(k,j)} \mid ( \mathcal{G}^{(k)},\eta_j )$ , for every $\eta_j \in \mathcal{H}$, and let 
\begin{displaymath}
z_{\text{INC}, \eta_j}^{(k)}=\left(\mathcal{G}_{\text{INC}, \eta_j}^{(k)},Y_{\text{INC}, \eta_j}^{(k)}\right)  
\end{displaymath}
be the observed data entailed by $\eta_j$.
\item Compute 
\begin{equation}
\mathcal{L}^{(k)}(\eta_j)=\int_{\Theta}  \mathcal{L}\left(\hat{Q}\left(z_{\text{INC}, \eta_j}^{(k)}\right),Q(\theta)\right)p(\theta \mid z_{\text{INC}, \eta_j}^{(k)})  d\theta,\nonumber
\end{equation}
% \label{Eq:ExpLossMC}
which is a Monte Carlo version of Equation \ref{Eq:ExpLoss}. Assume that $\hat{Q}(\cdot)$ is the Bayes rule corresponding to $\mathcal{L}$ (see Section 4.2.1 of \cite{Rober}).
\item Compute
\begin{equation}
\widehat{ \mathcal{L}}(\eta_j)=\frac{1}{K} \sum_{k=1}^{K} \mathcal{L}^{(k)}(\eta_j),\nonumber
\end{equation}
%\label{Eq:LossDesMC}
which is the Monte Carlo version of Equation \ref{Eq:ExpLossDes}.
\item The optimal design is given by 
\begin{equation}
\eta^{\star} =\arg\min_{\eta \in \mathcal{H}}\widehat{ \mathcal{L}}(\eta).\nonumber
\end{equation}
\end{enumerate}

Let $\eta^\star$ be the optimal design and let $\eta^\circ$ be the design calibrated according to convention, \emph{i.e.}, the design such that the tuning parameters are set as the default values used by practitioners in publications. Our first series of simulation studies will be based on the following scheme:
\begin{description}
\item[1a] Generate $\left( \mathcal{G}^{(k)}, Y^{(k)}\right)$ for  $k \in \left\{1,2,\dots,K \right\}$.
\item[2a] Given $\mathcal{G}^{(k)}$, simulate $I^{(k)} \mid ( \mathcal{G}^{(k)},\eta^\circ )$ and $I^{(k)} \mid ( \mathcal{G}^{(k)},\eta^\star)$, and let 
\begin{displaymath}
z_{\text{INC}, \eta^\circ}^{(k)}=\left(\mathcal{G}_{\text{INC}, \eta^\circ}^{(k)},Y_{\text{INC}, \eta^\circ}^{(k)}\right)  \quad \text{and} \quad 
z_{\text{INC}, \eta^\star}^{(k)}=\left(\mathcal{G}_{\text{INC}, \eta^\star}^{(k)},Y_{\text{INC}, \eta^\star}^{(k)}\right).
\end{displaymath}
\item[3a] Compute $\mathcal{L}^{(k)}(\eta^\circ)$ and $\mathcal{L}^{(k)}(\eta^\star)$.
%=\int_{\Theta}  \mathcal{L}\left(\hat{Q}\left(z_{\text{INC}, \eta^\circ}^{(k)}\right),Q(\theta)\right)p(\theta \mid z_{\text{INC}, \eta^\circ}^{(k)})  d\theta
%\end{equation}
%and
%\begin{equation}\label{Eq:ELDefault}
% \mathcal{L}^{(k)}(\eta^\star)=\int_{\Theta}  \mathcal{L}\left(\hat{Q}\left( z_{\text{INC}, \eta^\star}^{(k)}  \right),Q(\theta)\right)p(\theta \mid z_{\text{INC}, \eta^\star}^{(k)})  d\theta.
%\end{equation}
\item[4a] Compute
\begin{equation}   
\widehat{ \mathcal{L}}(\eta^\star)=\frac{1}{K} \sum_{k=1}^{K} \mathcal{L}^{(k)}(\eta^\star) \quad \text{and} \quad \widehat{ \mathcal{L}}(\eta^\circ)=\frac{1}{K} \sum_{k=1}^{K} \mathcal{L}^{(k)}(\eta^\circ).\nonumber
\end{equation}
%\label{Eq:ELDesOpt}
%and
%\begin{equation}\label{Eq:ELDesDef}
%\widehat{ \mathcal{L}}(\eta^\circ)=\frac{1}{K} \sum_{k=1}^{K} \mathcal{L}^{(k)}(\eta^\circ).
%\end{equation}
\end{description}
Different designs will be compared based on descriptive summaries of these quantities.

%We now proceed to discuss different summaries of simulations conducted this way. The objective of such summaries is to provide insights of how much is gained in terms of expected loss.

%Summaries
%\begin{itemize}
%\item Compare $\widehat{ \mathcal{L}}(\eta^\star)$ \emph{vs.} $\widehat{ \mathcal{L}}(\eta^\circ)$ using a table.
%\item Compare the distributions of $\widehat{ \mathcal{L}}^{(n)}(\eta^\star)$ and $\widehat{ \mathcal{L}}^{(n)}(\eta^\circ)$ in the same plot. 
%\item Plot the distribution of
%\begin{displaymath}
%\frac{\widehat{ \mathcal{L}}^{(n)}(\eta^\star)}{\widehat{ \mathcal{L}}^{(n)}(\eta^\circ)}.
%\end{displaymath}
%\end{itemize}

We also want to get better understanding of the average improvement gained by using a Bayesian experimental design framework when $\theta$ is specified. To achieve this, we performed simulations based on the idea of risk. This requires the following modifications:
\begin{description}
%\item For each $z^{(k)}=(\mathcal{G}^{(k)},Y^{(k)})$, sample
%\begin{displaymath}
%I^{(k)} \mid ( \mathcal{G}^{(k)},\eta^\circ ) \quad \text{and} \quad I^{(k)} \mid ( \mathcal{G}^{(k)},\eta^\star) . 
%\end{displaymath}
%\item Obtain
%\begin{displaymath}
%z_{\text{INC}, \eta^\circ}^{(k)}=\left(\mathcal{G}_{\text{INC}, \eta^\circ}^{(k)},Y_{\text{INC}, %\eta^\circ}^{(k)}\right)  \quad \text{and} \quad 
%z_{\text{INC}, \eta^\star}^{(k)}=\left(\mathcal{G}_{\text{INC}, \eta^\star}^{(k)},Y_{\text{INC}, %\eta^\star}^{(k)}\right).
%\end{displaymath}
\item[3b] Compute
\begin{displaymath}
 \mathcal{L}_{\mathcal{R}}^{(k)}(\eta^\circ)=\mathcal{L}(\hat{Q}\left(z_{\text{INC}, \eta^\circ}^{(k)}  \right),Q(\theta))\quad\text{and}\quad \mathcal{L}_{\mathcal{R}}^{(k)}(\eta^\star)=\mathcal{L}(\hat{Q}\left(z_{\text{INC}, \eta^\star}^{(k)}  \right),Q(\theta)).
\end{displaymath}
%and
%\begin{displaymath}
% \mathcal{L}_{\mathcal{R}}^{(k)}(\eta^\star)=\mathcal{L}(\hat{Q}\left(z_{\text{INC}, \eta^\star}^{(k)}  \right),Q(\theta)).
%\end{displaymath}
\item[4b] Compute
\begin{equation}\label{Eq:RiskD}
\widehat{ \mathcal{R}}(\eta^\circ) = \frac{1}{K} \sum_{k=1}^{K} \mathcal{L}_{\mathcal{R}}^{(k)}(\eta^\circ)\quad \text{and}, \quad
\widehat{ \mathcal{R}}(\eta^\star) = \frac{1}{K} \sum_{k=1}^{K} \mathcal{L}_{\mathcal{R}}^{(k)}(\eta^\star),\nonumber
\end{equation}
which are Monte Carlo estimates of the frequentist risk.
%and
%\begin{equation}\label{Eq:RiskOptimal}
%\widehat{ \mathcal{R}}(\eta^\star)=\frac{1}{K} \sum_{k=1}^{K} \mathcal{L}_{\mathcal{R}}^{(k)}(\eta^\star).
%\end{equation}
\end{description}
We follow a similar strategy to compare designs from an information theoretic perspective.

\begin{description}
%\item For each $z^{(k)}=(\mathcal{G}^{(k)},Y^{(k)})$, sample
%\begin{displaymath}
%I^{(k)} \mid ( \mathcal{G}^{(k)},\eta^\circ ) \quad \text{and} \quad I^{(k)} \mid ( %\mathcal{G}^{(k)},\eta^\star) . 
%\end{displaymath}
%\item Obtain
%\begin{displaymath}
%z_{\text{INC}, \eta^\circ}^{(k)}=\left(\mathcal{G}_{\text{INC}, \eta^\circ}^{(k)},Y_{\text{INC}, \eta^\circ}^{(k)}\right)  \quad \text{and} \quad 
%z_{\text{INC}, \eta^\star}^{(k)}=\left(\mathcal{G}_{\text{INC}, \eta^\star}^{(k)},Y_{\text{INC}, \eta^\star}^{(k)}\right).
%\end{displaymath}
\item[3c] Compute
\begin{displaymath}
p(Z \mid z_{\text{INC}, \eta^\circ}^{(k)})\quad p(Z \mid z_{\text{INC}, \eta^\star}^{(k)}).
\end{displaymath}
%and
%\begin{displaymath}
% \mathcal{L}_{\mathcal{R}}^{(k)}(\eta^\star)=\mathcal{L}(\hat{Q}\left(z_{\text{INC}, \eta^\star}^{(k)}  \right),Q(\theta)).
%\end{displaymath}
\item[4c] Compute
\begin{eqnarray}\label{Eq:RiskD}
\widehat{ \psi}(\eta^\circ)& = & \frac{1}{K} \sum_{k=1}^{K} H(p(Z \mid \theta^*), p(Z \mid z_{\text{INC}, \eta^\circ}^{(k)}))\quad \text{and}\nonumber\\
\widehat{ \psi}(\eta^\star)& = &\frac{1}{K} \sum_{k=1}^{K} H(p(Z \mid \theta^*), p(Z \mid z_{\text{INC}, \eta^\star}^{(k)}))\nonumber,
\end{eqnarray}
which are Monte Carlo estimates of the score proposed in Equation \ref{Eq:InfoScore}.
%and
%\begin{equation}\label{Eq:RiskOptimal}
%\widehat{ \mathcal{R}}(\eta^\star)=\frac{1}{K} \sum_{k=1}^{K} \mathcal{L}_{\mathcal{R}}^{(k)}(\eta^\star).
%\end{equation}
\end{description}
%Summaries
%\begin{itemize}
%\item Compare $\widehat{ \mathcal{R}}(\eta^\star)$ \emph{vs.} $\widehat{ \mathcal{R}}(\eta^\circ)$ using a table.
%\item Compare the distributions of $\widehat{ \mathcal{R}}^{(n)}(\eta^\star)$ and $\widehat{ \mathcal{R}}^{(n)}(\eta^\circ)$ in the same plot. 
%\item Plot the distribution of
%\begin{displaymath}
%\frac{\widehat{ \mathcal{R}}^{(n)}(\eta^\star)}{\widehat{ \mathcal{R}}^{(n)}(\eta^\circ)}.
%\end{displaymath}
%\end{itemize}

\subsection{Empirical Results for the Decision Theory Approach}
\addtocounter{example}{-5}
\begin{example}[continued] In this example, the number of referrals (denoted by $m$) is optimised for RDS.  We considered the mean squared error and the mean posterior loss as optimisation criteria. The comparison was performed for 3 different choices of prior (Table \ref{Tab:ResCoup}). The number of seeds ($w_0$) was set as 5. In all simulations, we observed that $m=3$ was optimal for the 3 choices of the prior. The differences in terms of posterior loss were clearer for priors corresponding to higher density of the network.
\end{example}

%\begin{table}[t!]
%\begin{center}
%\begin{tabular}{|l | c | c | c | c |}
%\hline
%Size & Density & $m$ & $\log(\text{loss})$ & $\log(\text{MSE})$ \\
%\hline
% $200$ &  $\frac{1}{200}$          &  2 &   -3.387 $\pm$ 0.0317    &   -1.382 $\pm$ 0.0734  \\
% $200$ & $\frac{1}{200}$         &  3 &     -3.391 $\pm$ 0.0288    &   -1.385 $\pm$ 0.0858   \\
% $200$ &$\frac{1}{200}$          &  4 &     -3.385 $\pm$ 0.0285   &  -1.381  $\pm$ 0.0663  \\
% $200$ &$\frac{1}{200}$        &  5 &       -3.376 $\pm$ 0.0300   &   -1.378 $\pm$ 0.0727\\
% $200$ &$\frac{1}{200}$        &  6 &       -3.376 $\pm$ 0.0295   &    -1.378 $\pm$ 0.0661 \\
%\hline
%\hline
% $200$ &  $\frac{1}{99}$          &   2     &    -3.306 $\pm$ 0.0184 &  -1.385 $\pm$ 0.0790     \\
% $200$ & $\frac{1}{99}$         &   3     &      -3.321 $\pm$ 0.0213  & -1.415 $\pm$ 0.0853      \\
% $200$ &$\frac{1}{99}$          &   4     &      -3.318 $\pm$ 0.0210  & -1.404 $\pm$ 0.0852     \\
% $200$ &$\frac{1}{99}$        &   5     &        -3.311 $\pm$ 0.0197  & -1.392 $\pm$ 0.0732   \\
% $200$ &$\frac{1}{99}$        &   6     &        -3.304 $\pm$ 0.0185  & -1.383 $\pm$ 0.0624    \\
% \hline
% \hline
%\end{tabular}
%\caption{Log expected loss and log mean square error for RDS designs for different priors and choices for $m$. %An Erd\"{o}s-R\'{e}nyi model was assumed and the sample size was set to $50$.}\label{Tab:ResCoup}
%\end{center}
%\end{table}
%\end{example}

\begin{table}[t!]
\begin{center}
\begin{tabular}{| c | c | c |}
\hline
 $m$ & $\log(\text{loss})$ & $\log(\text{MSE})$ \\
\hline
    2 &   -3.387 $\pm$ 0.0317    &   -1.382 $\pm$ 0.0734  \\
    3 &     -3.391 $\pm$ 0.0288    &   -1.385 $\pm$ 0.0858   \\
    4 &     -3.385 $\pm$ 0.0285   &  -1.381  $\pm$ 0.0663  \\
    5 &       -3.376 $\pm$ 0.0300   &   -1.378 $\pm$ 0.0727\\
    6 &       -3.376 $\pm$ 0.0295   &    -1.378 $\pm$ 0.0661 \\
\hline
\hline
% $200$ &  $\frac{1}{99}$          &   2     &    -3.306 $\pm$ 0.0184 &  -1.385 $\pm$ 0.0790     \\
% $200$ & $\frac{1}{99}$         &   3     &      -3.321 $\pm$ 0.0213  & -1.415 $\pm$ 0.0853      \\
% $200$ &$\frac{1}{99}$          &   4     &      -3.318 $\pm$ 0.0210  & -1.404 $\pm$ 0.0852     \\
% $200$ &$\frac{1}{99}$        &   5     &        -3.311 $\pm$ 0.0197  & -1.392 $\pm$ 0.0732   \\
% $200$ &$\frac{1}{99}$        &   6     &        -3.304 $\pm$ 0.0185  & -1.383 $\pm$ 0.0624    \\
% \hline
% \hline
\end{tabular}
\caption{Log expected loss and log mean square error for RDS designs for different priors and choices for $m$. An Erd\"{o}s-R\'{e}nyi model was assumed and the sample size was set to $50$, the size of the underlying graph as $200$ and the density of the graph as $\frac{1}{200}$.}\label{Tab:ResCoup}
\end{center}
\end{table}

\begin{example}[continued] In this example, the number of seeds ($w_0$) was optimised for RDS. We considered the mean squared error and the mean posterior loss as optimisation criteria. The comparison was performed for 3 different choices of prior (Table \ref{Tab:ResSeed}). The number referrals ($m$) was set as 3. In all simulations, we observed that more seeds lead to a reduction in the posterior loss, being 15 the optimal in all scenarios.

\begin{table}[t!]
\begin{center}
\begin{tabular}{| c | c | c |}
\hline
 Number of seeds & $\log(\text{loss})$ & $\log(\text{MSE})$ \\
\hline
   5     &    -3.351 $\pm$ 0.0337 &   -2.3279 $\pm$ 0.0196   \\
   7     &     -3.384 $\pm$ 0.0347  &  -2.4472 $\pm$ 0.0342     \\
   10     &     -3.396 $\pm$ 0.0338 &   -2.6265 $\pm$ 0.0434   \\
   12     &       -3.386 $\pm$ 0.0377 &   -2.7616 $\pm$ 0.0206  \\
   15     &       -4.005 $\pm$ 0.0370 &   -2.9208 $\pm$ 0.0290    \\
 \hline
\hline
%$\frac{1}{99}$          &   5     &    -3.775 $\pm$ 0.0334 &   -2.7958$\pm$ 0.0333   \\
%$\frac{1}{99}$         &   7     &      -3.832 $\pm$ 0.0371 &  -2.8538$\pm$  0.0430    \\
%$\frac{1}{99}$          &   10     &      -3.911 $\pm$ 0.0353 &  -2.9586 $\pm$ 0.0347   \\
%$\frac{1}{99}$        &    12     &       -3.951 $\pm$ 0.0381 &  -2.9586$\pm$ 0.0370 \\
%$\frac{1}{99}$        &   15     &        -4.096 $\pm$  0.0329 &  -3.0457$\pm$  0.0387   \\
% \hline
% \hline
% $200$ &  $\frac{1}{50}$          &   5     &    -3.865 $\pm$ 0.0396 &   -2.8860 $\pm$ 0.0224   \\
% $200$ &  $\frac{1}{50}$         &   7     &      -3.942 $\pm$ 0.0388 &  -2.9208 $\pm$  0.0426    \\
% $200$ &  $\frac{1}{50}$          &   10     &      -4.385 $\pm$ 0.0359 &  -3.0457$\pm$ 0.0237   \\
% $200$ &  $\frac{1}{50}$        &    12     &       -4.592 $\pm$ 0.0344 &  -3.0457 $\pm$ 0.0248 \\
% $200$ &  $\frac{1}{50}$        &   15     &        -4.783 $\pm$  0.0376 &  -3.0969$\pm$  0.0207   \\
% \hline
%\hline
\end{tabular}
\caption{Expected loss for RDS designs for different priors and choices for the change point of the design. An Erd\"{o}s-R\'{e}nyi model was assumed and the sample size was set to $50$, the size of the underlying graph as $200$ and the density of the graph as $\frac{1}{200}$.}\label{Tab:ResSeed}
\end{center}
\end{table}

\end{example}

\begin{example}[continued] In this example, the number of referrals ($m$) in RDS is allowed to vary in time. This is done by making this number a function of wave; this function is shaped by a parameter $\eta$. We optimised according to the mean squared error and the mean posterior loss. The comparison was performed for 3 different choices of prior (Table \ref{Tab:ResOneP}). The number of seeds $(w_0)$ was set as 5. In all simulations, we observed that $\eta=1.5$ was optimal for the 3 choices of the prior.

\begin{table}[t!]
\begin{center}
\begin{tabular}{| c | c | c |}
\hline
Exponent & $\log(\text{loss})$ & $\log(\text{MSE})$ \\
\hline
  2     &    -3.036 $\pm$  0.0281&   -3.148  $\pm$ 0.0450  \\
 1.5     &   -3.187 $\pm$  0.0287&  -3.221  $\pm$ 0.0382    \\
  1     &      -3.022 $\pm$  0.0275&  -2.920  $\pm$  0.0457 \\
  0.5     &     -3.070  $\pm$  0.0250&  -3.130  $\pm$ 0.0362 \\
  $\frac{1}{3}$     &    -2.920 $\pm$  0.0283 &  -3.154 $\pm$ 0.0397    \\
 \hline
\hline
% $\frac{1}{99}$          &   2     &    -3.187 $\pm$ 0.0288  &   -2.958  $\pm$ 0.0364  \\
% $\frac{1}{99}$         &   1.5     &   -3.327 $\pm$ 0.0283 &   -3.096   $\pm$ 0.0372  \\
% $\frac{1}{99}$          &   1     &      -3.251 $\pm$ 0.0284 &   -2.886   $\pm$  0.0473\\
% $\frac{1}{99}$        &   0.5     &     -3.236 $\pm$ 0.0284 &   -2.795   $\pm$ 0.0455\\
% $\frac{1}{99}$        &   $\frac{1}{3}$     &    -3.214 $\pm$ 0.0277 &  -2.823   $\pm$ 0.0402   \\
% \hline
% \hline
% $200$ &  $\frac{1}{50}$          &   2     &    -3.187 $\pm$ 0.0282  &   -2.920  $\pm$ 0.0443  \\
% $200$ & $\frac{1}{50}$         &   1.5     &   -3.327 $\pm$ 0.0286 &   -3.000   $\pm$ 0.0465  \\
% $200$ &$\frac{1}{50}$          &   1     &      -3.251 $\pm$ 0.0268 &   -2.853   $\pm$  0.0376\\
% $200$ &$\frac{1}{50}$        &   0.5     &     -3.236 $\pm$ 0.0264 &   -2.853   $\pm$ 0.0432\\
% $200$ &$\frac{1}{50}$        &   $\frac{1}{3}$     &    -3.214 $\pm$ 0.0251 &  -2.823   $\pm$ 0.0437   \\
% \hline
%\hline
\end{tabular}
\caption{Expected loss for RDS designs for different priors and choices for the parameter of the design. An Erd\"{o}s-R\'{e}nyi model was assumed and the sample size was set to $50$, the size of the underlying graph as $200$ and the density of the graph as $\frac{1}{200}$.}\label{Tab:ResOneP}
\end{center}
\end{table}

\end{example}

\begin{example}[continued] In this example, we allow the number of referrals ($m$) in RDS to take one of two values $(\lambda_H,\lambda_L)$ in each wave and once the bigger value is picked, that parameter of the design remains constant. In a sense this is similar to Example \ref{Ex:Curve}.  We computed the same summaries as in Examples \ref{Ex:NCoupons} - \ref{EX:Seeds}, this is, we first compared the expected loss (Table \ref{Tab:ResTwoS}) for different priors.  The number of seeds ($w_0$) was set as 5. For two of the simulations we observed that setting the breakpoint equal to 3 was the optimal, for the simulation corresponding to the denser graphs, we observed that 2 for the breakpoint was the optimal.

\begin{table}[t!]
\begin{center}
\begin{tabular}{| c | c | c |}
\hline
 Change Point & $\log(\text{loss})$ & $\log(\text{MSE})$ \\
\hline
  1     &    -3.376 $\pm$ 0.0400 &   -3.3665 $\pm$ 0.0441   \\
  2     &     -3.387 $\pm$ 0.0360  &  -3.4466 $\pm$ 0.0370     \\
  3     &     -3.396 $\pm$ 0.0377 &   -3.8206 $\pm$ 0.0472   \\
  4     &       -3.376 $\pm$ 0.0375 &   -3.5461 $\pm$ 0.0409  \\
  5     &       -3.366 $\pm$ 0.0392 &   -3.4420 $\pm$ 0.0401    \\
 \hline
%\hline
% $\frac{1}{99}$          &   1     &    -3.366 $\pm$ 0.0393 &   -2.7212 $\pm$ 0.0223   \\
% $\frac{1}{99}$         &   2     &      -3.408 $\pm$ 0.0366 &  -2.8239 $\pm$  0.0239    \\
% $\frac{1}{99}$          &   3     &      -3.431 $\pm$ 0.0380 &  -2.9586 $\pm$ 0.0192   \\
% $\frac{1}{99}$        &    4     &       -3.387 $\pm$ 0.0378 &  -2.8860  $\pm$ 0.0224 \\
% $\frac{1}{99}$        &   5     &        -3.376 $\pm$  0.0386 &  -2.7695 $\pm$  0.0195   \\
% \hline
 %\hline
 %$200$ &  $\frac{1}{50}$          &   1     &    -3.382 $\pm$ 0.0387 &   -2.8239 $\pm$ 0.0212   \\
 %$200$ &  $\frac{1}{50}$         &   2     &      -3.422 $\pm$ 0.0396 &  -2.9586 $\pm$  0.0218    \\
 %$200$ &  $\frac{1}{50}$          &   3     &      -3.545 $\pm$ 0.0369 &  -2.9208 $\pm$ 0.0201   \\
 %$200$ &  $\frac{1}{50}$        &    4     &       -3.392 $\pm$ 0.0389 &  -2.8860  $\pm$ 0.0198 \\
 %$200$ &  $\frac{1}{50}$        &   5     &        -3.381 $\pm$  0.0382 &  -2.8239 $\pm$  0.0199   \\
 %\hline
\hline
\end{tabular}
\caption{Expected loss for RDS designs for different priors and choices for the change point of the design. An Erd\"{o}s-R\'{e}nyi model was assumed and the sample size was set to $50$, the size of the underlying graph as $200$ and the density of the graph as $\frac{1}{200}$.}\label{Tab:ResTwoS}
\end{center}
\end{table}

\end{example}

\begin{example}[continued] In this example, we optimised the pair $(m,w_0)$. This was done by using the dynamic programming formulation: The first decision involving $w_0$, then, based on the reported degrees for the seeds, a second decision is made regarding $m$. We computed the same summaries as in Examples \ref{Ex:NCoupons} - \ref{EX:Switch}, this is, we first compared the expected loss (Table \ref{Tab:ResTwoS}) for different priors.  %For (?) scenarios, the pair (?) was optimal according to the squared (multilinear, intrinsic) loss. 

\begin{table}[t!]
\begin{center}
\begin{tabular}{| c | c | c |}
\hline
 $(m,w_0)$ & $\log(\text{loss})$ & $\log(\text{MSE})$ \\
\hline
  $(3,5)$     &    -3.352 $\pm$ 0.0331 &   -2.3276 $\pm$ 0.0197   \\
  $(3,10)$     &   -3.398 $\pm$ 0.0336 &   -2.6267 $\pm$ 0.0432   \\
  $(3,15)$     &   -4.025 $\pm$ 0.0371 &   -2.9211 $\pm$ 0.0287    \\
  $(4,5)$     &    -3.329 $\pm$ 0.0287 &   -1.3811  $\pm$ 0.0651  \\
  $(4,10)$     &    -3.252 $\pm$ 0.0333 &   -2.6284 $\pm$ 0.0483   \\
  $(4,15)$     &    -4.088 $\pm$ 0.0391 &   -2.9232 $\pm$ 0.0293    \\
  $(5,5)$     &     -3.312 $\pm$ 0.0300   &   -1.378 $\pm$ 0.0727\\
  $(5,10)$     &    -3.214 $\pm$ 0.0341 &   -2.6645 $\pm$ 0.0464   \\
  $(5,15)$     &    -4.025 $\pm$ 0.0368 &   -2.9211 $\pm$ 0.0302    \\
 \hline
\hline
\end{tabular}
\caption{Expected loss for RDS designs for different priors and choices for the pair $(m,w_0)$. An Erd\"{o}s-R\'{e}nyi model was assumed and the sample size was set to $50$, the size of the underlying graph as $200$ and the density of the graph as $\frac{1}{200}$.}\label{Tab:Seeds_and_Ref}
\end{center}
\end{table}

\end{example}

\subsection{Empirical Results for the Data Compression Approach}

%\begin{center}
%\begin{figure}[h!]
%\begin{center}
%\advance\leftskip-3cm
%\advance\rightskip-3cm
%\includegraphics[scale=0.8]{SBM1.png}
%\caption{Upper: The lower the value of the mean of the Hellinger distance distribution the better for a sampling designs to be recruited regarding $K$, which is the number of the communities. Middle: The lower the value of the mean of the expected squared loss distribution of the predictive distribution, $\mathbb{E}_{K}(K-\hat{K})^2$, the better for a sampling designs to be recruited regarding community number. Down: The lower the value of the mean of the expected absolute loss distribution of the predictive distribution, $\mathbb{E}_{K}(\mid K-\hat{K} \mid)$, the better for a sampling designs to be recruited regarding communities.}
%\label{Fig:Simulation1}
%\end{center}
%\end{figure}
%\end{center}

To explore the performance of our method, we compared the rankings implied by it to the rankings implied by the decision theory approach.  As the underlying model for the network, we assumed a SBM with  $K=10$ blocks and probabilities of inclusion, $\alpha_{i,i}=0.9$, and $\alpha_{i,j}=0.1$, for $i\neq j$. $N$ was set as $100$ for all the simulation regimes. All posterior samples for random network parameters were obtained via Markov chain Monte Carlo with a burn-in of 5000 samples and 1000 posterior samples. \\

We present the results in Table \ref{Tab:Seeds_and_Ref} and the figures in the supplementary material. Both simulation studies present empirical evidence why our approach could resembles decision theory methods (\cite{Lindley} and \cite{ChaloVerd}). It also provides evidence that our method produces robust rankings of sampling designs; here, the robustness is with respect to the choice of the loss function. We considered loss functions associated to point estimation and prediction.

%\begin{figure}[h!]

%\advance\leftskip-3cm
%\advance\rightskip-3cm
%\includegraphics[scale=0.9]{LSM.png}
%\caption{Upper: The lower the value of the mean of the Hellinger distance distribution the better for a sampling designs to be recruited regarding $\alpha$, which is the regression coefficient. Middle: The lower the value of the mean of the expected squared loss distribution of the predictive distribution, $\mathbb{E}_{\alpha}(\alpha-\hat{\alpha})^2$, the better for a sampling designs to be recruited regarding the regression coefficient. Down: The lower the value of the mean of the expected absolute loss distribution of the predictive distribution, $\mathbb{E}_{\alpha}(\mid \alpha-\hat{\alpha} \mid)$, the better for a sampling designs to be recruited regarding the regression coefficient.}
% \label{Fig:Simulation2}

%\end{figure}

\begin{table}
\begin{center}
\begin{tabular}{ |p{1.9cm}|p{1cm}|p{1.5cm}|p{1.5cm}|p{1.1cm}|p{1.1cm}| }
 \hline
 $I\mid\mathcal{G}$ & DC & SL (P) & AL (P) & SL & AL \\
 \hline
S (2,2,2)& 0.443  & 82.917  &9.238& 79.297&9.137\\
S (3,3,2)& 0.293 &   37.081  & 6.153 & 36.028& 6.253\\
S (3,2,3) & 0.112& 16.721 & 4.065  & 15.729&4.153\\
 \hline
RDS (2,2,2)&0.275 & 33.291  & 5.812  & 32.876  & 5.974\\
RDS (3,3,2)&0.219 & 20.917  & 5.249 &20.385  & 5.298\\
RDS (3,2,3)&0.197  & 21.021  & 5.397 & 21.746  & 5.464\\
 \hline

%LSM&$\alpha$&S (2,2,2)& 0.398   & 70.002  &8.085&68.682  &7.901\\
%LSM&$\alpha$&S (3,3,2)& 0.246  &   32.997  & 5.064 & 31.827  & 5.003\\
%LSM&$\alpha$&S (3,2,3) & 0.133 & 13.723 & 3.032  &  12.829 & 3.237\\
% \hline
%LSM&$\alpha$& RDS (2,2,2)&0.238 & 31.769  & 5.902  & 30.076  & 5.902 \\
%LSM&$\alpha$&RDS (3,3,2)&0.152 & 18.177  & 3.979 &16.994  & 3.007\\
%LSM&$\alpha$&RDS (3,2,3)&0.140  & 18.922  & 4.092 & 17.375  & 3.394\\
% \hline
\end{tabular}
\caption{ Rankings for different sampling designs $I\mid\mathcal{G}$, for the SBM. Here, the feature of interest $\tau(\mathcal{G})$ are the number of communities $K$, which corresponds to the number of blocks.
The rankings are obtained via the data compression approach (DC) and the decision theory approach via point estimation and prediction (P). For point estimation and point prediction, we used the square loss (SL) and the absolute loss (AL).}\label{Tab:Simulations}

%Means of Hellinger Distances Distribution (MHD) and means of Predictive Posterior (P.P), for point prediction, and Posterior (P.) Quadratic and Absolute Mean Distribution (MSE and MAE), for point estimation, for six different sampling designs in the settings of number of communities in SBM ($K$) and regression coefficient in latent space model ($\alpha$).
  \end{center}
\end{table}

\section{Discussion}
As far as we know, our methodology is the first one to apply tools from Bayesian experimental design to the problem of sampling on a social network. This allows a systematic calibration of the tuning parameters of different designs, given a full probability model for all sources of uncertainty and priors. The sources of uncertainty we considered were: the unobserved part of the network, the variability regarding the sampling mechanism and the uncertainty associated with the parameters of the model.\\

The challenges for this problem were, mainly: to incorporate all relevant sources of uncertainty, to deal with non-ignorable sampling designs and to write computer code that could be run in parallel. The first two challenges were addressed using the framework proposed in \cite{LunagAirol}. \\

One of our main contributions is that we provide results for understanding how different formalisms for comparing sampling designs relate in the context of network data. The criterion based on sufficiency and the Goel-DeGroot Criterion were discussed. It is worth mentioning that the main limitation of the Goel-DeGroot Criterion discussed in this paper is a consequence of the ignorability of some designs.

As the results show, the choices of tuning parameters matter, in the sense that there can lead to substantial differences in the loss or MSE. It was interesting to observe that the relationship  between the value of the tuning parameter and the MSE can be non-trivial (i.e., non-monotone).\\

Not surprisingly, the results sensitive to prior. In this case prior encodes the density of the network, since an Erd\"{o}s-R\'{e}nyi model was assumed. For networks with lower density, choices of the tuning parameter tend to have less impact on the expected loss (or MSE) when compared with networks with higher density. \\

Future work includes: To develop methodology for sequential problems, where learning from the network topology will assist the sampling, and propose adaptive designs, where not only learning from the topology, but learning from the responses can inform future decisions regarding the sampling. We also plan to apply this methodology for the case where degrees are observed (for the sampled nodes) with noise. This would involve dealing with issues as non-ignorable coarsening.

%\section*{Acknowledgement}

\bibliographystyle{biometrika}
\bibliography{mybib1}

\clearpage
\newpage
\appendix
%\appendixone
\section*{Appendix 1}
Proof of Theorem \ref{Th:DesignSuff1}
\begin{proof}

We start by defining an indexing $\varphi(\cdot)$ of the nodes contained in $\mathcal{V}_\text{INC}$. This indexing is such that:
\begin{enumerate}
\item if nodes $(i,j)$ were collected during waves $(w_x,w_y)$, respectively, whith $w_x<w_y$, then $\varphi(i)<\varphi(j)$
\item if nodes $(i,j)$ were collected during the same wave $w_x>1$, then $\varphi(i)<\varphi(j)$ if $(i,j)$ were referred by nodes $(h,k)$, respectively, with $\varphi(h)<\varphi(k)$.
\item Nodes corresponding to seeds are indexed according to a random permitation of $\left\{ 1,2,\dots, s \right\}$.
\item if nodes $(i,j)$ belong to the same connected component as the seeds $(h,k)$, respectively, with $\varphi(k)<\varphi(h)$, then $\varphi(i)<\varphi(j)$.
\end{enumerate}

1. By applying $\varphi(\cdot)$ into $\mathcal{V}_\text{INC}$, the adjacency matrix implied by $\mathcal{G}_\text{INC}$ will have a block structure. By making the rows corresponding to wave $w+1$ of this adjacency matrix equal to zero, we obtain a matrix with the same block structure as the one implied from a realisation of $I_{\text{LT}(s,r,w)}$; clearly, this mapping is a projection. Even more, the probability of observing a realisation from $I_{\text{LT}(s,r,w+1)}$, and then applying the indexing and the projection mapping, is the same as the probability from observing a realisation from $I_{\text{LT}(s,r,w)}$. This follows from marginalising the observations corresponding to the last wave from the generative model for link-tracing. \\

2. As in the previous case, apply the indexing $\varphi(\cdot)$ and then a projection mapping that make the entries of the adjacency matrix associated to the connected component containing $s+1$ equal to zero.  Analogous to the previous case, this is equivalent to marginalising the observations corresponding to the $(s+1)$ seed from the generative model for link-tracing. \\

3. We will prove by induction on the number of waves. For $w=1$, the result is trivial, since, instead of choosing $r+1$ neighbours at random from each seed, by applying a projection on the block structure, we obtain a realisation from a design where $r$ neighbours are chosen at random. We call this operation \emph{thinning}. Computing the marginalisation confirms this. For $w=2$, the result also follows. We can proceed by thinning the second wave first, then the edges from the first wave. Once the edges from the first wave have been thinned, the nodes recruited during the first wave incident to the deleted edges are removed from $\mathcal{G}_\text{INC}$, as well as the edges from the second wave incident to those nodes. Again, the result can be verified via marginalisation. The next step is to apply the induction hypothesis \emph{i.e.} the result is valid for $w=k$ waves. Finally, we prove the result is valid for $w=k+1$: first we perform thinning on the $k+1$ wave, then, thanks to the hypothesis of induction, the result holds for the first $k$ waves. After applying the corresponding thinning to the first $k$ waves, the nodes recruited during the $k=th$ wave incident to the deleted edges are removed from $\mathcal{G}_\text{INC}$, as well as the edges from the $(k+1)-th$ wave incident to those nodes.\\

4, 5 and 6 follow from the following observation: RDS can be represented as the adjacency matrix implied by link tracing and a vector storing the degrees. Both of these folloing the same indexing for the vertex set. Here, we define the corresponding projection mappings by extending the ones we defined for 1,2 and 3 so that, when edges are removed from the matrix, the contents of the entry of the vector corresponding to the node with highest indexing is also removed.

\end{proof}
 
Proof of Proposition \ref{Prop:DesignSuff0}

\begin{proof}

1. Composition of the theorem 1 gives: $I_{\text{LT}(s+a,r+b,w+c)}\succeq I_{\text{LT}(s,r,w)}$ with $a,b,c \succeq 0$ integers. E.g. inductively we have $\dots \succeq I_{\text{LT}(s+2,r,w)} \succeq I_{\text{LT}(s+1,r,w)}\succeq I_{\text{LT}(s,r,w)}$ from the first relationship of theorem 1 and combining the first with the second relationships of theorem 1 $I_{\text{LT}(s+1,r,w)}\succeq I_{\text{LT}(s,r,w)}$ and $I_{\text{LT}(s+1,r+1,w)}\succeq I_{\text{LT}(s+1,r,w)}$ we have $I_{\text{LT}(s+1,r+1,w)}\succeq I_{\text{LT}(s,r,w)}$.\\

2. The same logic of composition of theorem 1 part 4, 5, 6 gives: $I_{\text{RDS}(s+a,r+b,w+c)}\succeq I_{\text{RDS}(s,r,w)}$ with $a,b,c  \in \mathbb{Z}_{>0}$.\\
\end{proof}

Proof of Theorem \ref{Th:DesignSuff2}

\begin{proof}

To construct the counter-example, we assume that the underlying graph is regular and large enough so it does not exhaust the designs.\\

Mapping (3) decreases the expected degree of the observed subgraph, while the mapping used in (2) does not decrease nor increase the expected degree and the mapping (1) may decrease the expected degree. Combined, these statements imply that a composition of these mappings can only decrease the expected degree.\\

Mapping (2) decreases the number of connected components, while (1) and (3) do not affect the number of connected components. Combined, these statements imply that a composition of these mappings can only decrease the number of connected components.\\

Therefore, a relationship of sufficiency cannot be stablished between $I_{\text{LT}(s,r,w)}$ and $I_{\text{LT}(s+1,r-1,w)}$, since $I_{\text{LT}(s+1,r-1,w)}$ will have more connected components and lower expected degree when compared to $I_{\text{LT}(s,r,w)}$.\\

\end{proof}

Proof of Theorem \ref{Th:Monotone}

\begin{proof}

The notion proposed in Section \ref{Sec:DesignCompress} is given by:

\begin{eqnarray}\label{Eq:InfoMarios}
 \mathbb{E}_X(\text{KL}(p(\theta), p(\theta \mid \mathcal{G})))& = & \mathbb{E}_X\left\{-\sum p(\theta) \log\left( \frac{ p(\theta\mid \mathcal{G})}{p(\theta)}\right)\right\} \nonumber \\
                                     & = & \mathbb{E}_X\left\{-\sum p(\theta) \log\left( \text{Lik}(\mathcal{G};\theta) \right)\right\}\nonumber \\
                                     & = & \mathbb{E}_X\left\{- \mathbb{E}_\theta\left[ \log\left( \text{Lik}(\mathcal{G};\theta) \right) \right]\right\}.
\end{eqnarray}
Let $p(\mathcal{G})$ and $p_\ast(\mathcal{G})$ denote the predictive distribution implied by the priors $p(\theta)$ and $p_\ast(\theta)$, respectively. The notion presented in Definition \ref{Def:DeGroot} can be written as:
\begin{eqnarray}\label{Eq:IAmDeGroot}
J(p(\theta),p_\ast(\theta)) & = & \text{KL}(p(\mathcal{G}),p_\ast(\mathcal{G}))\nonumber \\
    & = & -\sum p(\mathcal{G}) \log \left[  \frac{p_\ast(\mathcal{G})}{p(\mathcal{G})} \right]\nonumber \\
    & = & -\sum p(\mathcal{G}) \log \left[  p_\ast(\mathcal{G}) \right] + \sum p(\mathcal{G}) \log \left[  p(\mathcal{G}) \right]\nonumber \\
    & = & -\mathbb{E}_X\left[ \log\left( p_\ast(\mathcal{G}) \right) \right]+\mathbb{E}_X\left[ \log\left( p(\mathcal{G}) \right) \right] \nonumber\\
    & = & -\mathbb{E}_X\left[ \log\left(  \sum p(\mathcal{G}\mid \theta) p_\ast(\theta)   \right) \right]+
    \mathbb{E}_X\left[ \log\left(  \sum p(\mathcal{G}\mid \theta) p(\theta)   \right) \right] \nonumber\\
    & = & \mathbb{E}_X\left\{- \log\left( \mathbb{E}_{\theta_\ast}\left[ \text{Lik}(\mathcal{G};\theta) \right] \right) \right\}- \mathbb{E}_X\left\{- \log\left( \mathbb{E}_\theta\left[ \text{Lik}(\mathcal{G};\theta) \right] \right) \right\}.
\end{eqnarray}
It follows from Jensen's inequality:
\begin{displaymath}
- \log\left( \mathbb{E}_\theta\left[ \text{Lik}(\mathcal{G};\theta) \right] \right) \leq - \mathbb{E}_\theta\left[ \log\left( \text{Lik}(\mathcal{G};\theta) \right) \right]
\end{displaymath}
since these terms appear under positive and negative sign in Expressions \ref{Eq:InfoMarios} and \ref{Eq:IAmDeGroot}, respectively. It is not possible to establish a monotone relationship. In addition, Expression \ref{Eq:IAmDeGroot} contains an additional term, which is a function of $p_\ast(\theta)$; this is a piece of information independent from the other tem. These facts imply that it is not feasible to establish a monotone relationship between both notions.\\

%Is there a difference, in practical terms between both notions? One way to answer this is by finding  a pair of choices for $p(x_{\text{INC}}\mid x)$ (\emph{i.e.}, sampling designs) such that one criterium declares them as equal, while the other does not. 

\end{proof}

Proof of Proposition \ref{Prop:Entropy}
\begin{proof}
The Erd\"{o}s-R\'{e}nyi model can be represented as a vector of iid random variables with distribution $\text{Ber}(\alpha)$ with $\frac{N(N-1)}{2}$ entries.  Since the entropy of a random vector with independent entries is the sum of the entropy for the marginals (Section 1.2 of \cite{MezaMont}), the entropy for the Erd\"{o}s-R\'{e}nyi with parameters $(N,\alpha)$ is given by
\begin{displaymath}
H_{\text{ER}}=-\frac{N(N-1)}{2}\left[ \alpha \log(\alpha) + (1-\alpha) \log(1-\alpha)\right]
\end{displaymath}
The entropy for $\text{Ber}(\alpha)$ has $[0,1]$ as its range, reaching the maximum at $0.5$ (Example 1.6 \cite{MezaMont}). Theorem \ref{thm:Compress} implies that, to optimally compress the Erd\"{o}s-R\'{e}nyi model, it is necessary to use a vector with at least $\llcorner H \lrcorner$ entries; where such a vector may contain statistics based on the full graph. One consequence of this is that sparse graphs (and very dense graphs) are much easier to compress than graphs where $\alpha \approx 0.5$, where the optimally compressed adjacency matrix will have as many entries as the original.\\
A similar statement can be made for the SBM with fixed $K$ number of blocks. Let $\beta_i$ denote the probability of allocating a node to block $i\in\left\{ 1,\dots,K\right\}$ and $\alpha_{i,j}$ the probability of including an edge between a node allocated to block $i$ and a node allocated to node $j$. Therefore, the entropy associated to a single entry of the adjacency matrix is given by:
\begin{displaymath}
2\times H_{\text{Block}}+\tilde{H}_{\text{Inclusion}},
\end{displaymath}
where
\begin{eqnarray}
H_{\text{Block}}& = & -\sum_{i=1}^K \beta_i \log(\beta_i)\nonumber\\
\tilde{H}_{\text{Inclusion}} & = & -\sum_{\left\{i,j\right\}\in \left\{1,2,\dots,K \right\}} \beta_i\beta_j\left[  \alpha_{i,j}\log(\alpha_{i,j})+(1-\alpha_{i,j})\log(1-\alpha_{i,j}) \right]\nonumber.
\end{eqnarray}
In contrast with the Erd\"{o}s-R\'{e}nyi model, the entries of the adjacency matrix are not independent for the SBM. Still, it is straightforward to derive the entropy for the adjacency matrix:
\begin{equation}
H_{\text{SBM}}=2\times N \times H_{\text{Block}}+\binom{N}{2}\times\tilde{H}_{\text{Inclusion}}\nonumber
\end{equation}
\end{proof}

Proof of Proposition \ref{Prop:ComprVsDeGroot} 

\begin{proof}
In general, we have
\begin{eqnarray}
\log\left( \mathbb{E}_\theta\left[ \text{Lik}(\mathcal{G};\theta) \right] \right) & = & \log\left( \mathbb{E}_\theta\left[ \sum_{\mathcal{G}_{\text{INC}}\sim x} p(\mathcal{G}_{\text{INC}}\mid x) p(\mathcal{G}\mid \theta) \right] \right) \nonumber \\
   & = & \log\left( \sum_{\mathcal{G}_{\text{INC}}\sim x} p(\mathcal{G}_{\text{INC}}\mid \mathcal{G})  \mathbb{E}_\theta\left[p(\mathcal{G}\mid \theta) \right] \right). \nonumber
\end{eqnarray}
Let us assume an Erd\"{o}s-R\'{e}nyi model for $\mathcal{G}$. In this context $\theta$ denotes the probability of inclusion. Let us assume a $\text{Beta}(\tau_1,\tau_2)$ as prior for $\theta$, then
\begin{equation}   
   \mathbb{E}_\theta\left[p(\mathcal{G}\mid \theta) \right] = \frac{\text{number of edges}+1+\tau_1}{\binom{N}{2}+2+\tau_1+\tau_2}
\end{equation}
For snowball we have:
\begin{equation}\label{Eq:ContSnow}
\sum_{\mathcal{G}_{\text{INC}}\sim \mathcal{G}} p(\mathcal{G}_{\text{INC}}\mid \mathcal{G})  \mathbb{E}_\theta\left[p(\mathcal{G}\mid \theta) \right]= \Arrowvert \mathcal{G}_{\text{INC}}\sim \mathcal{G} \Arrowvert \times \mathbb{E}_\theta\left[p(\mathcal{G}\mid \theta) \right],
\end{equation}
while for ego-centric design we have
\begin{equation}\label{Eq:ContEgo}
\sum_{\mathcal{G}_{\text{INC}}\sim x} p(\mathcal{G}_{\text{INC}}\mid x)  \mathbb{E}_\theta\left[p(\mathcal{G}\mid \theta) \right]= \binom{|\mathcal{V}|}{S}\times \mathbb{E}_\theta\left[p(\mathcal{G}\mid \theta) \right].
\end{equation}
In other words, because both designs are ignorable, the probability of observing a specific sample is constant with respect to the unobserved part of $\mathcal{G}$.

When plugging either Expression \ref{Eq:ContSnow} or Expression \ref{Eq:ContEgo} into Equation \ref{Eq:IAmDeGroot}, we obtain.
\begin{equation}
I(p(\theta),p_\ast(\theta))  =\mathbb{E}_X\left\{ \frac{\text{number of edges}+1+\tau_1}{\binom{N}{2}+2+\tau_1+\tau_2} \right\} - \mathbb{E}_X\left\{\frac{\text{number of edges}+1+\tau_1^{\ast}}{\binom{N}{2}+2+\tau_1^{\ast}+\tau_2^{\ast}}\right\}
\end{equation}

which means that both designs are equally able to discriminate between the posterior predictives implied by different priors. This in turn implies that both designs perform equally well according to the criterion presented in Definition \ref{Def:DeGroot}. The notion proposed in Section \ref{Sec:DesignCompress} would provide a different answer, since terms do not cancel out. This is relevant when looking at Equation \ref{Eq:ContEgo}, since, according to the Goel-DeGroot criterion, we should not be able to distinguish between 
two Ego-centric designs with different sample sizes, in contrast, the criterium proposed in Section \ref{Sec:DesignCompress}, which would capitalize on the term
\begin{displaymath}
\log \left[  \binom{|\mathcal{V}|}{S}    \right]
\end{displaymath}
to discriminate between designs based on sample size. Observe that it is the fact that Ego-centric design and snowball are ignorable what weakens the usefulness of the criterion presented in Definition \ref{Def:DeGroot}.
\end{proof}

\section*{Appendix 2}
%\subsection*{General}
In this appendix, we present the backward induction algorithm for multi-stage decision problems. The original formulation can be found in Chapter 3 of \cite{Bell}. First, we establish some additional notation: Let $\varphi[s,i]$ be a possible \emph{history}, \emph{i.e.}, a sequence of decisions and observations that constitute a path from the root of the decision tree to a node associated to a decision at stage $s$. Let $z^{\varphi[s,j]}_{\varphi[s-1,i]}$ the data required to augment $\varphi[s-1,i]$ to $\varphi[s,j]$.

I. For stage S
   \begin{enumerate}
   \item Compute the loss function for all the leaves of the decision tree. Each leaf is associated to an inference $a\left(z_{\varphi[S,i]}\right)$ and a value for $Q(\theta)$; the corresponding loss is given by $ \mathcal{L}\left(a\left(z_{\varphi[S,i]}\right),Q(\theta)\right)$.
   \item Compute the expected loss for each $a\left(z_{\varphi[S,i]}\right)$. The expectation is taken with respect to the posterior for $\theta$ given $z_{\varphi[S,i]}$; this is
      \begin{displaymath}
       \mathbb{E}\left( \mathcal{L}\left(a\left(z_{\varphi[S,i]}\right),Q(\theta)\right) \mid z_{\varphi[S,i]} \right)=
       \int_{\Theta}  \mathcal{L}(a(z_{\varphi[S,i]}),Q(\theta))p(\theta \mid z_{\varphi[S,i]})  d\theta.
      \end{displaymath}
   \item Compute the optimal decision associated to $z_{\varphi[S,i]}$; this is given by
      \begin{displaymath}
                       a^{\star}(z_{\varphi[S,i]}) =\arg\min_{a \in \mathcal{A}}\mathbb{E}\left( \mathcal{L}\left(a\left(z_{\varphi[S,i]}\right),Q(\theta)\right) \mid z_{\varphi[S,i]} \right).
     \end{displaymath}
   \end{enumerate}

II. For stage $s \in \left\{S-1,\dots,1 \right\}$
   \begin{enumerate}
   \item Compute the value of the loss function associated to each pair of the form $\left(a\left(z_{\varphi[s-1,i]}\right) ,z^{\varphi[s,j]}_{\varphi[s-1,i]}\right)$. This is given by
   %Here $z^{\varphi[s,j]}_{\varphi[s-1,i]}$ is the last piece of information disclosed for this stage, therefore it plays the same role as $Q(\theta)$ for stage $S$. The loss associated to $\left(a\left(z_{\varphi[s-1,i]}\right) ,z^{\varphi[s,j]}_{\varphi[s-1,i]}\right)$ is given by 
   \begin{displaymath}
      \mathcal{L}\left(a\left(z_{\varphi[s-1,i]}\right),z^{\varphi[s,j]}_{\varphi[s-1,i]}\right)=\mathbb{E}\left( \mathcal{L}\left( a^{\star}(z_{\varphi[s,j]}) \right) \mid z_{\varphi[s,j]} \right).
    \end{displaymath}
   \item Compute the expected loss for each $a\left(z_{\varphi[s-1,i]}\right)$. The expectation is taken with respect to the predictive distribution for $z^{\varphi[s,j]}_{\varphi[s-1,i]}$ given $z_{\varphi[s-1,i]}$; this is
   \begin{displaymath}
       \mathbb{E}\left( \mathcal{L}\left(a\left(z_{\varphi[s-1,i]}\right),z^{\varphi[s,j]}_{\varphi[s-1,i]}\right) \mid z_{\varphi[s-1,i]} \right)=
       \int_{\Theta}  \mathcal{L}\left(a\left(z_{\varphi[s-1,i]}\right),z^{\varphi[s,j]}_{\varphi[s-1,i]}\right)p\left(z^{\varphi[s,j]}_{\varphi[s-1,i]} \mid z_{\varphi[s-1,i]}\right)  d\theta.
      \end{displaymath}
   \item Compute the optimal decision associated to $z_{\varphi[s-1,i]}$; this is given by
  \begin{displaymath}
     a^{\star}(z_{\varphi[s-1,i]}) =\arg\min_{a \in \mathcal{A}}\mathbb{E}\left( \mathcal{L}\left(a\left(z_{\varphi[s-1,i]}\right),z^{\varphi[s,j]}_{\varphi[s-1,i]}\right) \mid z_{\varphi[s-1,i]} \right).
   \end{displaymath}
   \item Move to stage $s-1$, or stop if $s=1$.
   \end{enumerate}

\end{document}